\documentclass[conference]{IEEEtran}
\usepackage{amsfonts,amsmath,amstext,amssymb}
\newtheorem{conj}{Conjecture}
\newtheorem{thm}{Theorem}

\newtheorem{cor}{Corollary}
\newtheorem{prop}{Proposition}
\def\R{\mathbb{R}}

\usepackage[pdftex]{graphicx}
\usepackage[caption=false,font=footnotesize]{subfig}
\usepackage{epstopdf}
\usepackage{color}

\usepackage{ifthen}

\newboolean{showcomments}
\setboolean{showcomments}{true}
\ifthenelse{\boolean{showcomments}}
{ \newcommand{\mynote}[3]{
    \fbox{\bfseries\sffamily\scriptsize#1}
    {\small$\blacktriangleright$\textsf{\emph{\color{#3}{#2}}}$\blacktriangleleft$}}}
{ \newcommand{\mynote}[3]{}}

\newcommand{\A}{\mathbb{A}}
\newcommand{\D}{{\,\mathrm{d}}}
\newcommand{\E}[1]{\mathbb{E}#1}
\newcommand{\PP}[1]{\mathbb{P}#1}

\newcommand{\squishlist}{
 \begin{list}{$\bullet$}
  { \setlength{\itemsep}{0pt}
     \setlength{\parsep}{3pt}
     \setlength{\topsep}{3pt}
     \setlength{\partopsep}{0pt}
     \setlength{\leftmargin}{1.5em}
     \setlength{\labelwidth}{1em}
     \setlength{\labelsep}{0.5em} } }

\newcommand{\squishlisttwo}{
 \begin{list}{$\bullet$}
  { \setlength{\itemsep}{0pt}
     \setlength{\parsep}{0pt}
    \setlength{\topsep}{0pt}
    \setlength{\partopsep}{0pt}
    \setlength{\leftmargin}{2em}
    \setlength{\labelwidth}{1.5em}
    \setlength{\labelsep}{0.5em} } }

\newcommand{\squishend}{
  \end{list}  }

\allowdisplaybreaks

\begin{document}

\setlength{\pdfpagewidth}{8.5in}

\setlength{\pdfpageheight}{11in}

\title{Spatial Interactions of Peers\\ and Performance of File Sharing Systems }

\author{\IEEEauthorblockN{Fran{\c c}ois Baccelli}
\IEEEauthorblockA{INRIA--\'ENS\\
France
}
\and
\IEEEauthorblockN{Fabien Mathieu}
\IEEEauthorblockA{INRIA--University Paris 7\\
France
}
\and
\IEEEauthorblockN{Ilkka Norros}
\IEEEauthorblockA{VTT\\
Finland
}
}

\maketitle

\begin{abstract}
We propose a new model for peer-to-peer networking which takes
the network bottlenecks into account beyond the access. This
model allows one to cope with key features of P2P networking
like degree or locality constraints or the fact that distant peers
often have a smaller rate than nearby peers. We show that
the spatial point process describing peers in their steady
state then exhibits an interesting repulsion phenomenon.
We analyze two asymptotic regimes of the peer-to-peer network:
the fluid regime and the hard--core regime. We get closed form
expressions for the mean (and in some cases the law) of the
peer latency and the download rate obtained by a peer as well
as for the spatial density of peers in the steady state of each
regime, as well as an accurate approximation that holds
for all regimes. The analytical results are based on a mix of
mathematical analysis and dimensional analysis and have important
design implications. The first of them is the existence of a
setting where the equilibrium mean latency is a decreasing
function of the load, a phenomenon that we call super-scalability. 
\end{abstract}

\section{Introduction}

Peer-to-peer (P2P) architectures have been widely used over the Internet in the last decade. The main feature of P2P is that it uses the available resources of participating end users. In the field of content distribution (file sharing, live or on-demand streaming), the P2P paradigm has been widely used to quickly deploy low-cost, scalable, decentralized architectures.
For instance, the ideas and success of BitTorrent~\cite{bittorrent} have shown that distributed file-sharing protocols can provide practically unbounded scalability of
performance. Although there are currently many other architectures that compete with P2P (dedicated Content Distribution Networks, Cloud-based solutions, \ldots),
P2P is still unchallenged with respect to its low-cost and scalability features, and remains a major actor in the field of content distribution.

The Achilles' heel of todays' P2P content distribution is the access
upload bandwidth, as even high-speed Internet access connections are
often asymmetric with a relatively low uplink capacity. Therefore,
most theoretical models of P2P content distribution presented so far
have been `traditional' in the sense of assuming a common, relatively
low access bandwidth, in particular concerning the upload direction,
which functions as the main performance bottleneck. However, in a near
future the deployment of very high speed access (e.g. FTTH) will challenge the
justification of this assumption. 
This raises the need of new P2P models that describe what
happens when the access is not necessarily the main/only bottleneck
and that allow one to better understand the fundamental
limitations of P2P.

\subsection{Contributions}

{\bf {A new model.}} The first contribution of the present paper is
the model presented in Section~\ref{sec:model}, which
features the following two key ingredients which were lacking
in previous models of the literature on P2P dynamics:
1) a spatial component thanks to which the topology of
the peer locations is used to determine their interactions
and their pairwise exchange throughput;
2) a networking component allowing one to represent the
capacity of the network elements as well as the transport
protocols used by the peers and to determine the actual
exchange throughput between them.

More precisely, we consider a scenario where peers randomly
appear in some metric space, typically the Euclidean plane
representing the physical distance, and download from their
neighbors with a throughput that may depend on some distance or RTT
(it can be the case for e.g. TCP transport).
The typical P2P application we have in mind is a BitTorrent-like
file-sharing system. However, the high abstraction level of
our model also allows for interpretations beyond this framework.
Using proper QoS requirements, it could be extended to any
kind of P2P content distribution services (like live
and on-demand streaming). The space could also be a representation
of the peers' interests, the position of a peer representing its
own centers of interest. In such a space, two close peers
share common interests, and therefore are likely to exchange
more data.\\

{\bf {A promising form of scalability.}}
The rationale that is usually brought forward to explain P2P
scalability is that the overall service capacity growths with the number of peers. This allows the system to reach an equilibrium point no matters how popular the service is.
This equilibrium
was first analytically studied in \cite{qiusrikant04}, under
the traditional assumption mentioned above
that the upload/download capacity is the bottleneck
determining the exchange throughput obtained by peers.
The model proposed in \cite{qiusrikant04} leads to
an equilibrium point which exhibits the expected scaling
property in that the service latency can be shown to remain
constant when the system load increases.
In our new model, the equilibrium
point may exhibit a stronger form of scalability than that in
\cite{qiusrikant04}, that we propose to call super-scalability,
where the service latency actually {\em decreases} with the
system load.\\

{\bf {Conditions for super-scalability to hold.}}
As we shall see in Sections \ref{sectoy} and \ref{sec:theory},
this super-scalability phenomenon is not difficult to understand
from a pure queuing theory or graph theory viewpoint. Roughly
speaking, super-scalability can be shown to hold in a queue
whenever the service rate of a typical customer scales like
the number of customers in the system (rather than like a
constant as in \cite{qiusrikant04}). Equivalently, it is not
difficult to see that it holds if the peer interaction graph
is complete at any given time.

However, in practice, the network cannot sustain arbitrary high rates. Also, interactions
between peers are limited by degree constraints and by the
requirement to select peer connections with good throughput.
Section \ref{sec:lim} combines our model together with an
abstract network model to determine the conditions on the 
peering rules, on the network capacity and on the transport
protocols for which the mathematical analysis makes sense
and for which the super-scalability property can possibly survive.\\

{\bf {The laws of super-scalability.}}
The paper also provides a full analytical quantification of the
system at the equilibrium point: in addition to the latency formula,
it also provides closed form expressions for e.g. the density of
peers present in the P2P overlay or the rate obtained by each peer, 
as functions of the peering rules and the network parameters.
These equilibrium laws, which take specific forms
for each type of transport protocol, are the main analytical contributions
of the paper. These are gathered in Section~\ref{sec:theory}
for the simplest scenarios and in 
Sections \ref{sec:generalrate} and \ref{sec:extend} for a few variants that can be built on our model: generic rate functions, auxiliary servers,
seeding behavior of users, access bottleneck condition, 
etc.

These laws have important P2P
implications. In particular, they allow one to determine
optimal tuning of the parameters of the P2P algorithms
e.g. the optimal peering degree or the best parameters
of the transport protocols to be used within this context.

One theoretically novel feature of our model is the proof of a {\em
repulsion} phenomenon which was empirically observed
in \cite{DBLP:conf/sigcomm/OttoSCBS11}:
as close peers get faster rates, they quit
the system earlier, so a node ``sees'' fewer peers in its
immediate vicinity than one would expect by considering the
spatial entrance distribution alone. 
All these results are validated through simulations in Section \ref{sec:simulations}. 

\subsection{Related Work}

Our main scenario 
is inspired by a BitTorrent-like file-sharing protocol. In BitTorrent \cite{bittorrent}, a file is segmented
into small chunks and each downloader (called \emph{leecher}) exchanges
chunks with its neighbors in a peer-to-peer overlay network. A peer may continue to distribute chunks after it has completed its own download (it is called a \emph{seeder} then).
Theoretical studies and 
modeling have already provided relatively good understanding of BitTorrent performance.

Qiu and Srikant \cite{qiusrikant04} analyzed the effectiveness of P2P
file-sharing with a simple dynamic system model, focusing on the
dynamics of leechers and seeders. Massoulié and Vojnovic
\cite{massoulievojnovic08} proposed an elegantly abstracted stochastic
chunk-level model of uncoordinated file-sharing. 
In the case of non-altruistic peers (who do not continue as seeders), their results indicated that if the system has high input rate and starts with a large and chunk-wise sufficiently balanced population, it may perform well very long times without any seeder. However, instability may be encountered in the form of the ``missing piece syndrome'' identified by Mathieu and Reynier \cite{mathieu:missing}, where one (and exactly one!) chunk keeps existing in very few copies while the peer population grows unboundedly. Hajek and Zhu \cite{hajekzhu10,zhuhajekarx11} proved that the syndrome is unavoidable, if the non-altruistic peers enter empty-handed and if the peer arrival rate is larger than the chunk upload rate offered by persistent seeders. On the other hand, they also proved that the system becomes stable for any input rate, if the peers have enough altruism to stay as seeders as long as it takes to upload one chunk. The missing piece syndrome can be avoided even in the case of non-altruistic peers by using more sophisticated download policies at the cost of somewhat increased download times, see \cite{reittuiwsos09,norreiei11,oguzanantnor-arxiv}. The above results were obtained in a homogeneous, potentially fully connected network model. The present paper introduces a much less trivial family of peer interaction models, focusing on a bandwidth-centered approach similar to the one proposed by Benbadis \emph{et al.} \cite{benbadis08playing}. To avoid excessive layers of complexity, we neglect chunk-level modeling in this phase, although realizing that meeting the rare chunk problem will modify and enrich the picture in future research.

The natural feature of large variation of transfer speeds in P2P
systems has been considered in a large number of papers. For example,
part of the peers can rely on cellular network access that is an order
of magnitude slower than fixed network access used by the other
part. Such scenarios differ however substantially from our model,
where the transfer speeds depend on pair-wise distances but not on the
nodes as such.

There are some earlier papers considering P2P systems in a spatial
framework. As an example, Susitaival {\em et
  al.}\ \cite{pannet-susitaival-06} assume that the peers are randomly
placed on a sphere, and compare nearest peer selection with random
peer selection in terms of resource usage proportional to
distance. However, the distance has no effect on transfer speed in
their model. Our paper seems to be the first where a peer's
downloading rate is a function of its distances to other peers.

\section{Super--scalability Toy Example}
\label{sectoy}

Consider a system in steady state where 
jobs arrive to get some service.
This system will be said to be super-scalable if the mean
job latency decreases when the arrival rate increases
and all other system parameters remain fixed.

In order to understand how super-scalability can arise,
we propose the following two toy examples: consider a system where
peers arrive and want to download some file of size $F$.
Peers arrive in the system with intensity $\lambda$
and leave the system as soon as their own download is completed.

In our first toy example,
the access upload bandwidth is
considered as the main bottleneck. If we neglect issues related
to data/chunk availability, and if $U$ is the typical upload bandwidth
of a peer, then it makes sense to assume that $U$ is also the
typical download throughput experienced by each peer. In particular,
in the steady state (if any), the mean latency $W$ and the average
number of peers $N$ should be such that
\begin{equation}
	W=\frac{F}{U}\text{ and }\
	N=\lambda W=\frac{\lambda F}{U}\text{ (Little's Law).}
\label{eq:bottleneck-1.0}
\end{equation}
Although very simple, \eqref{eq:bottleneck-1.0} contains a core property of
standard P2P systems: the mean latency is independent of the arrival rate.
This is the \emph{scalability} property, which is one of the main
motivations for using P2P.

Now, imagine a second toy example based on a complete shift of the bottleneck paradigm.
Let the main resource bottleneck be the (logical, directed)
links between nodes instead of the nodes themselves. We should
then consider the typical bandwidth $C$ \emph{from one peer to
another} as the key limitation. If each peer is connected to every
other one (the interaction graph is complete at any time),
then the equilibrium Equation \eqref{eq:bottleneck-1.0}
should be replaced by
\begin{equation*}
	W=\frac{F}{(N-1)C}\text{ and }\
	N=\lambda W\text{, which leads to}\\
\label{eq:bottleneck-2.0}
\end{equation*}
\begin{equation*}
N=\sqrt{\frac{\lambda F}{C}+\frac{1}{4}}+\frac{1}{2}\text{ and }
W=\sqrt{\frac{F}{\lambda C}+\left(\frac{1}{2\lambda}\right)^2}+\frac{1}{2\lambda}\text{.}
\label{eq:bottleneck-2.1}
\end{equation*}
For $\frac{\lambda F}{C}\gg1$, this can be approximated by
\begin{equation}
N\approx\sqrt{\frac{\lambda F}{C}}\text{ and }\\
W\approx\sqrt{\frac{F}{\lambda C}}\text{.}\\
\label{eq:bottleneck-2.2}
\end{equation}
The behavior of this new system is quite different from the previous one.
Among other things, the service time is now inversely proportional
to the square root of the arrival intensity, so that
super-scalability holds.

In this toy example, the central reason for super-scalability
is rather obvious: the number of edges in a complete graph
is of the order of the square of the number of nodes,
and so is the overall service capacity.

The main question addressed in the present paper is
to better understand the fundamental limitations of P2P systems
and in particular to check whether super-scalability can possibly
hold in future, network-limited, P2P systems, where the throughput
between peers will be determined by transport protocols and  
network resource limitations rather than the upload capacity alone.
This requires the definition of a new model allowing one
to take both the latter and the former into account 
as well as the limitations inherent 
to P2P overlays like e.g. the constraints on the degree of the
peering graph, the availability of data/chunks, etc.

\section{Network Limited P2P Systems}
\label{sec:model}
\def\R{\mathbb{R}}

The aim of this section is to define a model meeting all the above requirements.

\subsection{Dynamics}

Our peers live in a spatial domain $D$. The domain can be some general Euclidean or even abstract metric space. It can describe physical distance between peers, distances derived from metrics in the underlying physical network, or even represent some semantic space.

For simplicity, we focus on a basic model where $D$ is the Euclidean plane
$\R^2$, but there is no basic difficulty in extending this framework. We also use sometimes an arbitrarily large torus as an approximation of $D$.

Assume that new peers arrive according to some time-space random process.
The set of the positions of peers present at time $t$ is denoted by $\Phi_t$.

Each peer ${p}$ has an individual service requirement $F_{p}>0$. In the basic
example where the service required by every peer consists of downloading one
and the same file, $F_{p}$ would most naturally be modeled as a constant $F$
describing the size of the file.

We assume that two peers at locations $x$ and $y$ serve each other at
rate $f(||x-y||)$, where $f$ is a non-negative function which we call
the {\em bit rate function} of the model\footnote{We implicitly assume that
bandwidth rates are automatically adjusted by the system, at the network layer, in a TCP-like fashion, or at the applicative layer, using a UDP-like approach.
}. This function describes the network
transport and connectivity limitations. We will see later how these limitations can be taken into account.

In order to focus on bandwidth aspects, we do not explicitly take into account issues related to chunk availability. Following the approach proposed by \cite{qiusrikant04}, we assume that filesharing effectiveness can be affected by some factor $\eta\leq 1$ because sometimes, a peer may not have any chunk that a neighbor would want. In the following, we omit $\eta$ by assuming that file sizes are always scaled by a factor $\frac{1}{\eta}$. We are aware, however, that handling chunk availability through a constant $\eta$ has some limitations, and we will point out the scenarios where chunk availability can become a real issue.
  
The services received from several peers are additive, so that the total 
download rate of a peer at $x$ is
$$
\mu(x,\phi_t)=\sum_{y\in\phi_t\setminus\{x\}}f(||x-y||).
$$

By symmetry, $\mu(x,\phi_t)$ is also the upload rate of a peer 
at $x$. In order for the access not to be a further limitation,
the access capacity of a peer at $x$ should exceed $\mu(x,\phi_t)$.
This is our default assumption here (access as a possible bottleneck is considered in
Section~\ref{sec:extend}).
 
A peer ${p}$ born at point $x_{{p}}$ at time
$t_{{p}}$ leaves the system when its service requirement has
been fulfilled, i.e.\ at time
$$
\tau_{{p}}
=\inf\{t>t_{{p}}:\,\int_{t_{{p}}}^t\mu(x_{{p}},\phi_s)ds\ge
F_{{p}}\}.
$$ 

A peer is usually called a \emph{leecher} if it has not completed its download, and \emph{seeder} if it has. Although this paper is mainly focused on leechers-only system, the situation where peers continue as seeders after having completed their service will be considered in Section~\ref{sec:extend}.

\subsection{Examples of Bit Rate Functions}\label{secebrf}

We will consider two basic cases throughout the paper:
\begin{enumerate}
\item peers use a TCP-like congestion control mechanism;
\item peers use UDP.
\end{enumerate}
In P2P, UDP is often used
in place of TCP. However, P2P-over-UDP protocols try to be
TCP--friendly \cite{utp,ledbat}: they are designed to respect
TCP flows and actually mimic TCP\footnote{For instance, TFRC
(www.ietf.org/rfc/rfc3448.txt) recommends that UDP
flows use the square root formula to predict the transfer rate that a TCP
flow would get and use this rate for throttling their traffic.  
The TCP model is hence directly applicable to such a setting.}.

Consider first the case where peers use TCP Reno. On the path between
two peers, let $\vartheta$ denote the packet loss probability and $\mathrm
{RTT}$ denote the round trip time.  Then the square root formula
\cite{ott} stipulates that the rate obtained on this path is
$ \frac{\xi}{{\mathrm {RTT}} \sqrt{\vartheta}},$
with $ \xi= \sim1.309$. Assuming the RTT to be proportional to
distance $r$ yields a transfer rate of the form
\begin{equation}
\label{eqr0}
g(r)=\frac{C}{r}.
\end{equation}
We can refine \eqref{eqr0} by assuming that $\mathrm {RTT}$ is not simply
linear in $r$ but some affine function of it, namely
$\mathrm{RTT}=ar+b$, where $a$ accounts for propagation delays in the
Internet path and $b$ accounts for the mean delay in the two access
networks. Then the transfer rate
between two peers with distance $r$ becomes
\begin{equation}\label{eqr1} g(r)=\frac C{r+q},\quad
\mbox{with}\quad
C=\frac{\xi}{a\sqrt{\vartheta}}, \quad q=\frac{b}{a}.
\end{equation}
Another natural model is that where one accounts for an overhead cost
of $c$ bits per second.  The transfer rate between two peers at distance
$r$ is then
\begin{equation}
\label{eqr2} 
g(r)= \left(\frac{C}{r+q}-c\right)^{+}\text{, with $(.)^{+}:=\max(.,0)$.}
\end{equation}

In the case where peers use UDP, on the path between
two peers, the transfer rate is of the form
\begin{equation}
\label{eqr0U}
g(r)={C}\text{, where $C$ is a constant.}
\end{equation}

\subsection{Connectivity Limitation}

Having specified some transfer rate function $g$, we notice that a peer
cannot interact with all other peers of the overlay network:
it would result in a full mesh overlay, impossible to handle
for large networks. Therefore, peers usually limit their neighborhood, 
for instance by selecting only peers within a certain distance
and/or by limiting its total number of neighbors.
This constraint is even more meaningful in the wireless contex,
as it can correspond to some transmission range.
This leads to the following choices for the bit rate function:
\begin{itemize}
\item {\em {Constant Range}} model: take
$f(r)=g(r)1_{r\le R}$ ($R$ is called the {\em range}), so that
\begin{equation}
\label{eq:nonexp}
\mu(x,\Phi)= \sum_{x_i\in \Phi, x_i\ne x}  1_{||x_i-x||\le R} \,g(||x_i-x||)\text{,}
\end{equation}
where $g$ is one of the functions considered above.
\item
{\em {Constant Number of Nearest Peers}} model: take the $L$
closest peers as the set of communicating neighbors. This rule is
non-symmetric and difficult to deal with exactly. To begin with, computing the effective rate $f$ between to peers at $x$ and $y$ is not a function of $||x-y||$ only, but of the configuration $\phi_t$.
\end{itemize}

In this paper, the {\em main model} will be that where the transfer rate
between two communicating peers is given by \eqref{eqr0} or \eqref{eqr0U}
and where the range is constant. More
general rate functions (e.g. as defined in \eqref{eqr1} and \eqref{eqr2})
and an approximation of connectivity defined by the number of peers will
be analyzed in Section~\ref{sec:extend}.

\begin{table}
\centering
{\scriptsize
\begin{tabular}{|l|l|l|}
\hline
Name & Description & Units \\
\hline
\hline
$C_{TCP}$ & Speed parameter & $bits\cdot s^{-1}\cdot m$ \\
\hline
$C_{UDP}$ & Speed parameter & $bits\cdot s^{-1}$ \\
\hline
$F$ & Mean file size & $bits$ \\
\hline
$R$ & Peering range & $m$\\
\hline
$\lambda$ & Leecher arrival rate & $m^{-2}\cdot s^{-1}$\\
\hline
$W$ & Mean latency & $s$\\
\hline
$\mu$ & Mean rate & $bits\cdot s^{-1}$\\
\hline
$U$ & Upload bottleneck & $bits\cdot s^{-1}$\\
\hline
\end{tabular}
}\caption{Table of Notation}
\vspace{-.3cm}
\end{table}

Let us stress again that the framework can be 
extended to more general metric spaces and/or
to more general rate functions. For instance,
in a noise limited wireless network of the 
Euclidean plane, it makes sense to assume
that the rate between two peers at distance $r$
is determined by some Signal to Noise Ratio
condition and is hence proportional to
$\log\left(1+\frac C {r^\alpha}\right)$ with
$\alpha>2$ the path loss exponent.
Of course the additive assumption on the
point-to-point rates only makes sense
in rather particular cases (e.g. orthogonal
channels) and more general models should
be considered within this wireless setting.
We will not pursue the general wireless setting in
the present paper. We will however
consider more general rate functions than the above TCP
and UDP functions in Section \ref{sec:extend} including
the above additive wireless setting,
which will be referred to as the SNR model.

\subsection{Mathematical Assumptions}

We assume that new peers arrive according to a
Poisson process with space-time intensity $\lambda$ ({\em
`Poisson rain'}). $\lambda$, expressed in $m^{-2}.s^{-1}$, 
describes the birth rate of peers: the number of peer arrivals taking place in
a domain of surface $A$ (expressed in $m^2$)
in an interval $[s,t]$ (in seconds) is a Poisson random
variable with parameter $\lambda A(t-s)$. 

For the sake of mathematical tractability,
we assume the $F_{p}$'s to be independent and identically distributed
random variables with finite expectation, denoted by
$F=\E(F_{p})$. More specifically, we assume in this paper that their
common distribution is exponential of mean $F$
in order to gain in mathematical tractability.

\begin{prop}
\label{lem0}
If the domain $D$ in which the peers live is compact, then 
$\phi_t$ is a Markov process which is
ergodic for any birth rate $\lambda>0$.
\end{prop}
The proof, which can be found in Appendix \ref{sapp0},
is based on a domination argument
which can easily be extended to unbounded domains. The existence of 
stationary regimes for $\phi_t$ in the case of an {\em unbounded
domain} then follows from this and a tightness argument.
However, the ergodicity of $\phi_t$ and the uniqueness
of its stationary regimes cannot be established as easily in
this case. Garcia and Kurtz
\cite{garciakurtz06} proved the existence and ergodicity of a wide
class of {\em attractive} spatial birth-and-death processes in infinite
domains. Extending their approach to our repulsive case (see below for
the terminology) seems feasible but goes way beyond what can be done
within the space limitations of the present paper. In what 
follows, for results stated on the (infinite) Euclidean plane case,
we conjecture that the spatial birth-and-death processes
of interest admit a unique stationary regime. In any case,
all our results can be rephrased on a large torus where
this conjecture is not needed.

\section{Mathematical Analysis}
\label{sec:theory}

In this section, we focus on the main model under TCP
\eqref{eqr0} with fixed range $R$. The results
on UDP \eqref{eqr0U} are provided as well.
We adopt the same strategy concerning
the proofs as above for Proposition \ref{lem0}: we give
proofs in the torus case, so as to provide the main ideas, 
but refrain from discussing their extensions to the
infinite Euclidean plane. The limiting arguments for
these extensions are left for future work. The final formulas
are however always given in the infinite Euclidean 
plane where they have a particularly simple form.

For the main model, the system has 4 basic parameters: 
the range $R$ in meters ($m$), the typical filesize $F$ in bits, the peer arrival rate
$\lambda$ in $m^{-2}\cdot s^{-1}$ and a rate constant $C$ in $bits\cdot m \cdot s^{-1}$ ($bits\cdot s^{-1}$ in the UDP case).

According to Proposition \ref{lem0},
the model admits a steady state regime where
the peers (in the basic model all leechers) form in $\R^2$ a
stationary and ergodic point process \cite{DalVJon:88}.

We denote by $\beta_o$ the density of the peer (leecher) point process, 
by $\mu_o$ the mean rate of a typical peer, by $W_o$ the mean latency
of a typical peer, and by $N_o$ the mean number of peers in a ball of radius $R$ around a typical peer, all in the steady state regime of the P2P dynamics.

In the following, we will also consider several approximations of the main model:
\begin{itemize}
	\item a {\em fluid regime/limit}, where the corresponding quantities will be denoted
by a $f$ subscript (e.g. $\beta_f$);
\item a {\em hard--core regime/limit} for
which we will use the notation $._{h}$ (e.g. $\beta_h$);
\item a heuristic description of the main model with a hat notation (e.g. $\hat{\beta}_0$)
\end{itemize}

In any of these regimes, Little's law tells that the average density verifies $\beta=\lambda W$. 

\subsection{Fluid Limit}
\label{subsec:simplefluid}

The fluid limit consists in assuming that the density is uniformly distributed in space at any time. In particular, in the fluid limit, the presence of one single peer in a given point does not impact the system.

From Campbell's formula \cite{DalVJon:88},
the mean total bit rate of a typical location 
of space (or equivalently of a newcomer peer) is
\begin{equation}
\label{meanrate}
\mu_f= \beta_f 2 \pi \int_{r= 0}^R (C/r) r dr=
\beta_f 2\pi C R.
\end{equation}

Now, the fluid limit hypothesis allows one
to assume that a peer sees $\mu_f$ during its whole lifetime.
We get that the mean latency of a peer is 
\begin{equation}
W_f= \frac F{\mu_f}.
\label{eq:wffmuf}
\end{equation}
Hence
\begin{equation}
\label{eq:equil}
\beta \mu= \lambda F.
\end{equation} 

From \eqref{meanrate}, \eqref{eq:wffmuf} and (\ref{eq:equil}), we have
\begin{equation}\beta_f= \sqrt{
\frac{\lambda F}{2 \pi CR}}, \ 
\mu_f = \sqrt{
\lambda F 2 \pi CR}, \ 
W_f = \sqrt{ \frac F{
\lambda 2 \pi CR}}.
\label{eq:cassimple}
\end{equation}
In the fluid limit, the mean number of peers in a ball of radius $R$ around a typical peer is
\begin{equation}
\label{eq:NN}
N_f = \pi R^2 \beta_f=\sqrt{\frac \pi 2} \sqrt{\frac{\lambda F R^3}{C}}\text{.}
\end{equation}

For UDP, the same reasoning gives:
\begin{eqnarray}
\beta_{f,{\mathrm{UDP}}} & = \sqrt{ \frac{\lambda F}{\pi CR^2}},  \quad
\mu_{f,{\mathrm{UDP}}} & = \sqrt{ \lambda F \pi CR^2}, \nonumber\\ 
W_{f,{\mathrm{UDP}}} & = \sqrt{ \frac F{ \lambda \pi CR^2}},\quad
N_{f,{\mathrm{UDP}}} & =\sqrt{\pi} \sqrt{\frac{\lambda F R^2}{C}}
\label{eq:cassimpleu}
\end{eqnarray}

As we see in the expression
for the mean latency in \eqref{eq:cassimple} and \eqref{eq:cassimpleu}
both the TCP and the UDP fluid limits exhibit 
the same super-scalability as the toy example: in spite
of the fact that the interactions are not as in the complete
graph and depend on the distance, the mean latency
decreases in $\frac 1 {\sqrt{\lambda}}$ when $\lambda$ tends to
infinity and everything else is fixed. 

\subsection{Dimensional Analysis}

At this point of the paper, the fluid limit is a thought experiment,
not necessarily related to the actual model. Dimensional 
analysis \cite{pi} helps to address this issue.

We first use the $\pi$-theorem~\cite{pi} to strip our problem from redundant variables: if we choose $R$ as a new distance unit, then the arrival intensity
becomes $l=\lambda R^2$, the download constant becomes $c=C/R$ and the other
parameters are unchanged. If we now define $F$ as an information unit,
then the download speed constant becomes $c=C/(RF)$ and the other parameters are unchanged.
Finally, if we take a time unit such that the download speed constant is $1$, we get a system
where all parameters are equal to $1$ but for the arrival rate which
is equal to $l=\frac{\lambda F R^3}{C}$. As the system itself is not affected by the choice of measurement units, all its properties only depend on the (dimensionless) parameter
\begin{equation}
\rho = \rho_{{\mathrm{TCP}}} = \frac{\lambda F R^3}{C}.
\label{eq:theorempi}
\end{equation}

The $\pi$-theorem allows some freedom in the choice of the parameter. By noticing that $N_f=\sqrt{\frac \pi 2} \sqrt {\rho}$, we can use $N_f$, which has a physical interpretation (the number of neighbors predicted by the fluid limit), instead of $\rho$.

By similar arguments, we have
\begin{equation}
\rho_{{\mathrm{UDP}}} = \frac{\lambda F R^2}{C}\text{,}
\label{eq:theorempiu}
\end{equation}
so we use $N_{f,{\mathrm{UDP}}} = \sqrt{\pi } \sqrt {\rho_{{\mathrm{UDP}}}}$.

The $\pi$-theorem tells that all systems that share the same
parameter $N_f$ are similar. Now consider the union of two independent systems that use the same parameters ($\lambda$, $F$, $C$, $R$): the real model, with latency $W_o$, and the fluid model, with latency $W_f$. The ratio $\frac{W_o}{W_f}$ is a dimensionless property of the overall system, therefore it is a function of $N_f$ only. 
In other words,
there exists a dimensionless function $M(N_f)$ such that:
\begin{equation}
W_o=M(N_f)W_f\text{.}
\label{eq:Wfull}
\end{equation}

From Little's law, we also deduce the density:
\begin{equation} 
\beta_o= \beta_f M(N_f)\text{.}
\label{eq:nonameinmind}
\end{equation}

These equations are true for both the TCP and UDP rates (with a different $M$ function in each case).

To summarize, although our system may be subject to complex interactions and is defined by four independent parameters, dimensional analysis allows one
to express its general behavior through a one-parameter function
$M$ (unknown), which expresses how far the real system is from its
fluid limit.

\subsection{Fluid as a Bound}

We now give a better understanding of the behavior of the real system through the following theorem.
\begin{thm}[Repulsion]\label{conjecture1} In the steady state,
\begin{equation}
\label{eq:palnopal}
 \E[ \sum_{x_i\in \Phi } f(||x_i||)] \ge \E_0[ \sum_{x_i\in \Phi \setminus\{0\}} f(||x_i||)],
\end{equation}
where $\E_0$ denotes expectation w.r.t.
$\PP_0$, the Palm probability \cite{DalVJon:88} w.r.t. the point process
$\Phi$.
\end{thm}
The proof is given in Appendix \ref{sapp1} in the torus case.
Theorem \ref{conjecture1} says that
there are less points (in terms of their $f$--weight)  in a ball 
of radius $R$ around a typical peer (i.e. under the Palm
probability) than in a ball of the same radius 
around a typical location of the Euclidean plane (i.e.
under the stationary probability $P$). This is what we call a repulsion effect.
\begin{cor} \label{cor1}
$M\geq 1$.
\end{cor}
\emph{Proof of corollary:}
Theorem \ref{conjecture1} is equivalent to saying that
$\mu_{o}\le \beta_o 2\pi CR.$
This, the relation
$W_{o}\ge F/\mu_{o}$ (which is obtained
by a direct convexity argument) and Little's law $\beta_o=\lambda W_o$
imply that
$\beta_o\ge \lambda \frac F {\beta_o 2\pi CR}$
which in turn implies 
$\beta_{o}\ge \beta_f$ and $M\geq 1$.

In other words, repulsion implies that the fluid regime is actually a lower (resp. upper) bound for the mean latency and the peer density (resp. the mean rate).
Now, the following theorem tells that the bound is tight.

\begin{thm}
\label{conjecture2}
When $N_f$ tends to infinity, $M$ tends to $1$, and the law of a typical peer latency converges weakly to an exponential random variable of parameter $W_f$.
\end{thm}
The sketch of proof is given in Appendix \ref{sapp2},
where it is shown that this regime is such that not only
the traffic is high but the peers also stay long enough
to make the fluctuations slow and weak.
By the almost constancy of the rate at any point, we get the almost 
exponentiality.

Theorem \ref{conjecture2} says that when the number of neighbors predicted by the fluid limit tends towards infinity, the system behaves like its fluid limit.

\subsection{Hard--Core Regime} 
\label{sec:hcr}

A stationary point process is {\em hard--core} with
exclusion radius $R$ if there is no pair of
points in the point process with a distance less than $R$.
\begin{conj}\label{conjecture3} 
When $N_f$ tends to 0, $N_fM(N_f)$ tends to 1, and 
the stationary peer point process tends to a hard--core point process
with exclusion radius $R$, with intensity $\beta_h$ and latency $W_h$ 
defined as follows:
\begin{equation}
\label{eq:hh}
\beta_h=\frac{1}{\pi R^2},\ 
W_h=\frac{1}{\lambda \pi R^2},\ 
\end{equation}
Moreover,
the cdf of the latency converges weakly to
\begin{equation}
1-\frac{e^{-\frac{t}{2{W_h}}}}{2},\quad t>0.
\label{eq:tdis-hardball}
\end{equation}
\end{conj}
Conjecture \ref{conjecture3} is supported by simulations (cf. Section \ref{sec:simulations}), and by the following insight on the hard--core behavior: when two peers are at distance $r\leq R$, the average time under TCP for one of them to disappear is less than $\frac{rF}{C}\leq \frac{RF}{C}$. If $N_f\ll 1$, \eqref{eq:cassimple}, \eqref{eq:NN} and Corollary \ref{cor1} give $\frac{RF}{C}\ll W_o$. In other words, when two peers are in range, one of them disappears almost instantly compared to the typical latency of the system, so when we take a snapshot of the system at a given time and finite area of space, it is likely that we see only peers out of range $R$ from each other.

A similar reasoning stands for UDP.

It is worthwhile mentioning that according to (\ref{eq:hh}),
the {\em volume fraction}
of the associated sphere packing model is 1/4 (since we have a density $\frac{1}{\pi R^2}$ of non-intersecting balls of radius $R/2$).
This volume fraction is hence the same as that of the Mat\'ern hard--ball model 
in the so called jamming regime (see e.g. \cite{FnT1}).

Let us stress that this hard-core regime is hardly desirable (performance largely below the one predicted by the fluid limit and extreme unfairness). Moreover, peer data exchanges are very sparse, so 
the fluid assumption on the exchange of chunks fails
to hold. Chunk availability becomes probably a bottleneck as important as bandwidth under these conditions, suggesting that the performance will be even worse should we take chunk exchanges explicitly into account.

For all these reasons, the hard--core regime, which we presented for completing the description of our model, should be avoided by all means. The discussions
on design will hence be in part focused on the
tuning of the system parameters that avoid this regime.

\subsection{Heuristic} 
\label{secheur}
For intermediate values of $N_f$, where fluid and hard--core limits do not apply, we propose a first order approximation.

For TCP, it consists in approximating $M$ by $\hat{M}$, the unique solution in $[1,\infty)$ of
\begin{equation}
\hat{M}^2\left(1-\frac{\hat M}{2N_f}\ln \left(1+\frac{2N_f}{\hat M}\right)\right)=1\text{.}
\label{eq:Mheuristic}
\end{equation}

In order to derive \eqref{eq:Mheuristic}, we use a heuristic factorization of the factorial
moment measure of order 3 \cite{DalVJon:88} which is described in Appendix \ref{rsappendheur}.
Informally, the method consists in computing an approximation
$\hat{u}_o$ of the  average rate of a peer assuming that: (i)  a
neighbor at distance $r$ from that peer  ``sees'' a rate
$\hat{u}_o+\frac{C}{r}$; (ii) in return, the peer ``sees'' at distance
$r$ a density of neighbors $\frac{\lambda F}{\hat{u}_o+\frac{C}{r}}$
(using \eqref{eq:equil}).

Under this approximation, the fluid equation \eqref{meanrate} now becomes
\begin{equation}
\begin{array}{rcl}	
\mu_o 
& \approx & \lambda F 2\pi C \int_0^R \frac 1 {\mu_o +\frac C r} dr\\
& = &  \lambda F 2\pi C  \left(\frac R {\mu_o} -\frac C {\mu_o ^2} \ln(1 + \frac{\mu_o R} C)\right),
\end{array}
\end{equation}
which leads to
\begin{equation}\label{eqhard}
\hat \mu_o^2 = \mu_f^2 \left(1- \frac {C} {\hat \mu_o R} \ln\left(1 + \frac{\hat \mu_o R} C\right)\right).
\end{equation}

Using $\hat{\mu}_0=\frac{\mu_f}{\hat{M}}$ and noticing that $\frac{\mu_f R}{C}=2N_f$, equation \eqref{eq:Mheuristic} follows.

This heuristic is in line with Theorem \ref{conjecture2}
and Conjecture \ref{conjecture3}.
When $N_f$ tends to $\infty$,
it follows from \eqref{eq:Mheuristic}  that
$\hat M\sim 1$.
This is in line with Theorem \ref{conjecture2}.
When $N_f$ tends to 0,
expanding the log in \eqref{eq:Mheuristic}  gives
$\hat M \sim \frac{1}{N_f}$,
which substantiates Conjecture \ref{conjecture3}.

In the UDP case, the same heuristic leads to
\begin{equation}
\hat M = \frac{1}{\hat M}+\frac{1}{N_f}\text{, so that}
\label{eq:Mheuu}
\end{equation}
\begin{equation}
\hat M=\sqrt{1+\left(\frac{1}{2N_f}\right)^2}+\frac{1}{2N_f},
\label{eq:Mheuusoluce}
\end{equation}
which also supports both Theorem \ref{conjecture2}
and Conjecture \ref{conjecture3}.
\subsection{Toy Example Revisited}

We revisit the example of Section \ref{sectoy}
within the more precise framework considered in
the present section (Poisson
arrivals, exponential file size). This toy example can be seen as
the UDP case on the torus, when the range
is large enough for all pairs of peers to be within
range. Assume the surface of the torus to be 1.
Then, geometry disappears and we have a birth and death process
for the total population with 
birth rate $\lambda$ and death rate in state $i$ equal to
$\mu(i)=\frac C F i(i-1).$
The state space is that of positive integers.
The local balance equations read
$$\pi(i-1)\lambda=\pi(i) \mu i(i-1),\quad i\ge 2.$$
The solution is
$$\pi(i)= \pi(1)\frac{1}{i!(i-1)!} \rho^{i-1},\quad i\ge 2,$$
where $\rho=\frac {\lambda F} C$.
Hence the mean is
\begin{eqnarray*}
\beta  = 
\frac{1+\sum_{i\ge 2} \frac{1}{((i-1)!)^2} \rho^{i-1}}
{1+\sum_{i\ge 2} \frac{1}{i!(i-1)!} \rho^{i-1}}
 = \sqrt{\rho} \frac{B(0,2 \sqrt{\rho})}{B(1,2 \sqrt{\rho})},
\end{eqnarray*}
with $B$ the BesselI function. 

In words, we have
\begin{equation}
\beta=\beta_f M(N_f)\text{with 
$\left\{\begin{array}{l}
\beta_f=N_f=\sqrt{\frac{\lambda F}{C}} \text{ and} \\
M(X)=\frac{B(0,2 X)}{B(1,2 X)}.
\end{array}\right.
$}
\label{eq:toyrevisited}
\end{equation}

We recognize the approximation \eqref{eq:bottleneck-2.2}, 
which implies super-scalability in $\frac{1}{\sqrt{\lambda}}$,
corrected with an $M$ function like in \eqref{eq:nonameinmind}.
We remark that for this toy example, we have the exact value of $M$
and not an asymptote or a heuristic, and we can verify
that Corollary \ref{cor1} and Theorem \ref{conjecture2} hold.

Notice also that if $\delta(i)=\mu i$, then
$\pi(i)=\frac{\rho^i}{i!} e^{-\rho}$ for all $i\ge 0$
so that 
$N=\rho$ and $W=1/\mu$. 
In this case, which in essence is that of \cite{qiusrikant04},
or equivalently that of the M/M/$\infty$ queue, where
we have scalability but no super-scalability.

\section{Simulation Results}
\label{sec:simulations}

In this section, we validate our results and substantiate our
results by means of simulations. For sake of computability, 
we approximate the infinite space by a torus of radius $1$.
For concision, we only present here the simulation results for the TCP case, but UDP results are completely similar.

As stated by the dimensional analysis, all systems can be described by the function $M$. The goal of the simulation is then to sample that function. We just have to fix three independent parameters and use the fourth one to run through all possible scenarios.

We decide to choose the following fixed parameter: $R=.1$, which gives a good trade-off between the torus as an approximation of the plan; $C=1$ (arbitrary choice); $W_f=100$. The last choice means that remaining free parameters are adjusted so that \eqref{eq:cassimple} values to $100$ in each experiment. That way, for all simulations, the fluid model will predict the same mean latency, so the measured latencies will give $M$ directly, up to a constant factor $W_f$.

We naturally use $N_f$ (defined by \eqref{eq:NN}) as the variable parameter.
We use $N_f$ instead of $\rho$ as main dimensionless parameter because it is strictly
equivalent from the point of view of dimensional analysis, yet it gives a direct meaning
to the variable (average number of neighbors in the fluid model).
The remaining input parameters of the system are then completely defined:
\begin{equation}\lambda=\frac{N_f}{\pi R^2 W_f}, \quad
F=\frac{2N_fCW_f}{R}.\label{eq:simu_parameters}
\end{equation}
We choose to use a discrete time simulator, with elementary time step set to $\tau=1$. With our settings, the resulting step transitions are empirically small enough for the discrete model to be a good approximation of the continuous model. In the end, we get a simulator that achieves the needed trade-off between speed and accuracy (an event-based simulator, for instance, would give exact rendering of the continuous model but would require a lot more of computation).
For each considered setting, the simulation runtime was adjusted so that about $20000$ peers could be observed in stationary state. All results presented are obtained through $10$ runs per setting.

\subsection{Properties of the $M$ Function}

\begin{figure}%
\begin{center}
\includegraphics[width=\columnwidth]{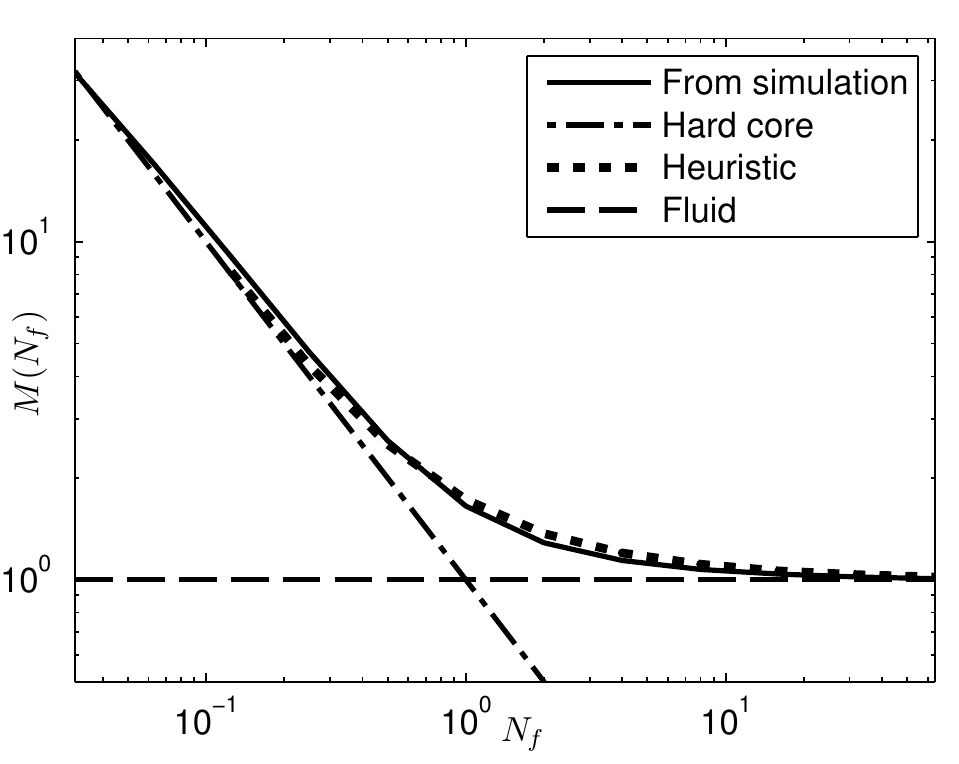}%
\caption{$M(N_f)$ in the TCP case.}%
\label{fig:nm}%
\end{center}
\end{figure}

We propose to start with a global study of the function $M$. We made simulations for $N_f$ varying from $1/32$ to $64$. Results are displayed Figure \ref{fig:nm}.

The empirical results are compared with
1) the fluid limit, $1$,
2) the hard--core limit, $\frac{1}{N_f}$, and 3) the heuristic formula (\ref{eq:Mheuristic}).

Figure \ref{fig:nm} allows us to check almost all results from previous section in one look: 
\begin{itemize}
\item
the fluid limit is a lower bound of the actual system (which is equivalent to Theorem \ref{conjecture1});
\item
as $N_f$ goes to $\infty$, the fluid bound becomes tight (this is Theorem \ref{conjecture2});
\item
as $N_f$ goes to $0$, the system behavior converges towards the hard--core limit (this is Conjecture \ref{conjecture3}).
\end{itemize}
Additionally, one checks that the heuristic \eqref{eq:Mheuristic} gives a
good approximation of $M$ for intermediate values of $N_f$,
while converging to the hard--core and fluid limits when $N_f$ goes to $0$ and $\infty$ respectively.

\subsection{Fluid Model}

We now propose to focus on the case $N_f=64$, in order to analyze the system in detail when it reaches the fluid limit. The value $M(64)$ given by simulations is $1.007$, which is higher than $1$ yet very close to it, as predicted by Theorem \ref{conjecture2}.

\begin{figure}%
\includegraphics[width=\columnwidth]{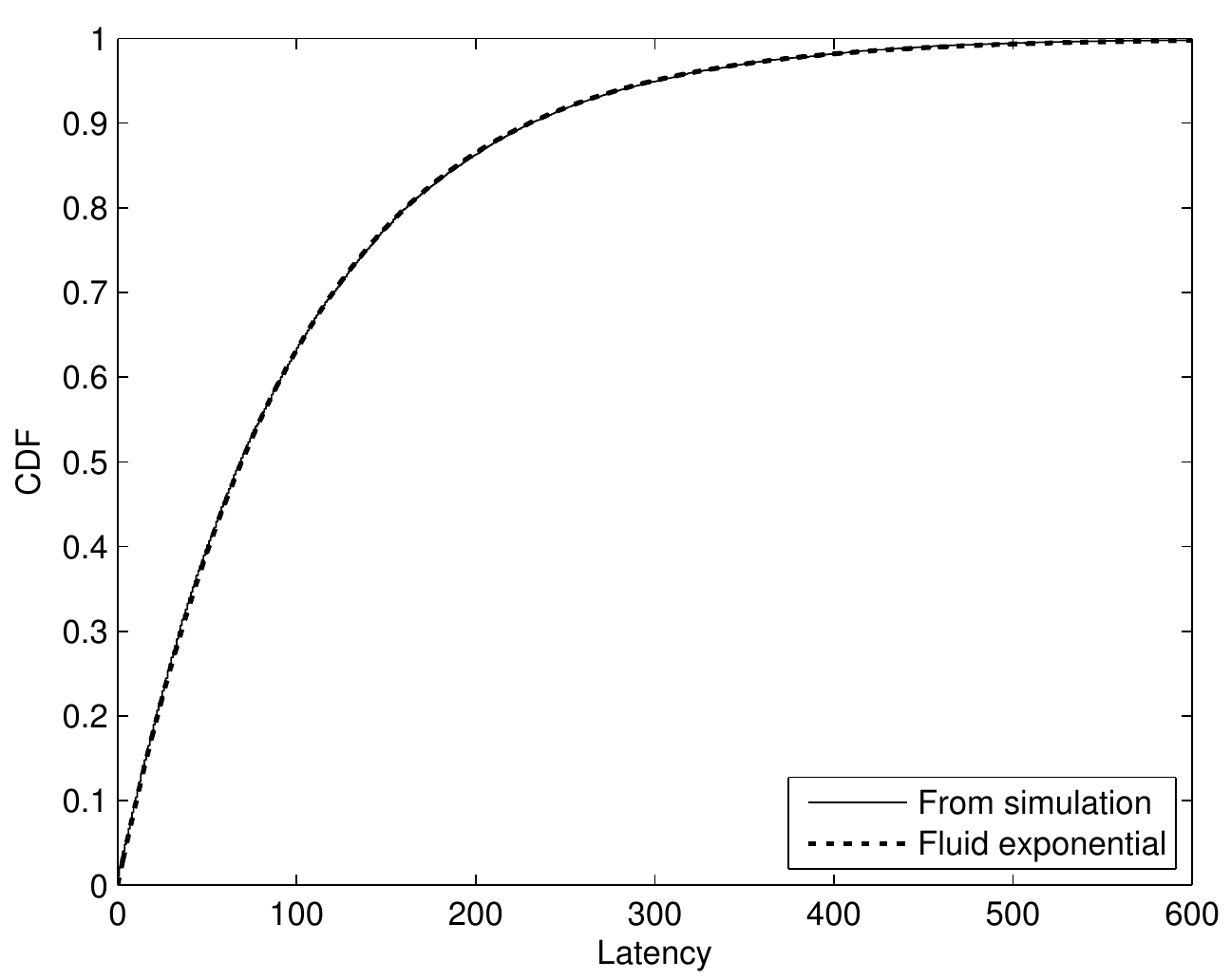}%
\caption{Cdf of latency for $N_f=64$.}%
\label{fig:cdf_latency_n64}%
\end{figure}

If one looks at the latency distribution, it is almost indistinguishable from an exponential distribution of mean $W_f$ (Figure \ref{fig:cdf_latency_n64}) as predicted by Theorem \ref{conjecture2}. 

In the fluid model, it is quite difficult to distinguish the system from a spatial birth and death process of birth parameter $\lambda$ and death parameter $1/W_f$, namely a Poisson point process of intensity $\beta_f$. Differences can only be spotted if small distances are involved. More precisely, two peers at distance $r$ have a mutual latency influence of $\frac{rF}{C}$, so one can expect Palm effect to become less visible when $\frac{rF}{C}$ is large enough compared to $W_f$. This allows us to show that $\frac{R}{N_f}$ is the critical distance below which the Palm effects become difficult to neglect. For $N_f=64$, this gives $\frac{R}{64}\approx 0.016$.

\begin{figure}%
\includegraphics[width=\columnwidth]{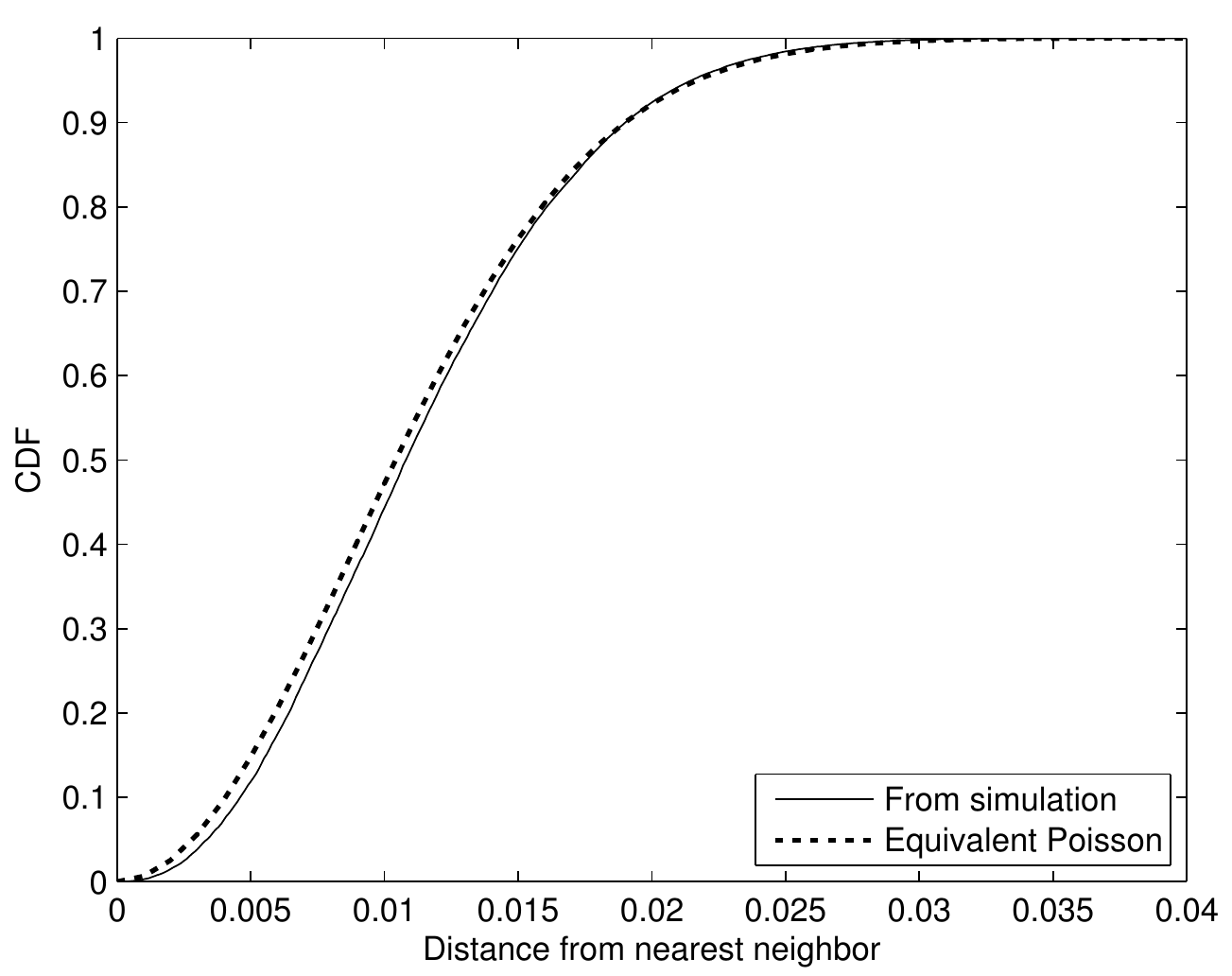}%
\caption{Cdf of nearest neighbor distance for $N_f=64$.}%
\label{fig:cdf_closest_n64}%
\end{figure}

In our case, the best way to differentiate the actual process from a Poisson process is to consider how far the closest neighbor of a peer is. While for a Poisson process the distance should be $\frac{1}{2\sqrt{\lambda W_f}}\approx 0.0111$ in average, simulation shows an actual average distance of $0.0115$: the nearest neighbor is slightly farther away by about $4\%$. If we go into detail by comparing the two distributions, it appears that the main gap appears for small distances (cf. Figure \ref{fig:cdf_closest_n64}), which supports the concept of critical Palm distance: if a peer gets a very close neighbor, both rates will be higher than usual, so one of them is likely to leave sooner, lowering the probability of finding very close neighbors in a random configuration. 
As $N_f$ tends towards $\infty$, we expect this difference to become negligible: the probability to get a neighbor so near that it will significantly affect the total rate becomes arbitrary low, so the repulsion effect becomes negligible.

\subsection{Hard--Core Model}

We conduct the same type of detailed study for $N_f=1/32$. For these parameters, the value $M(1/32)$ is now $31.6$, to compare with 
the hard--core model prediction $M_h=32$;
so the accuracy of the model is pretty good.

\begin{figure}%
\includegraphics[width=\columnwidth]{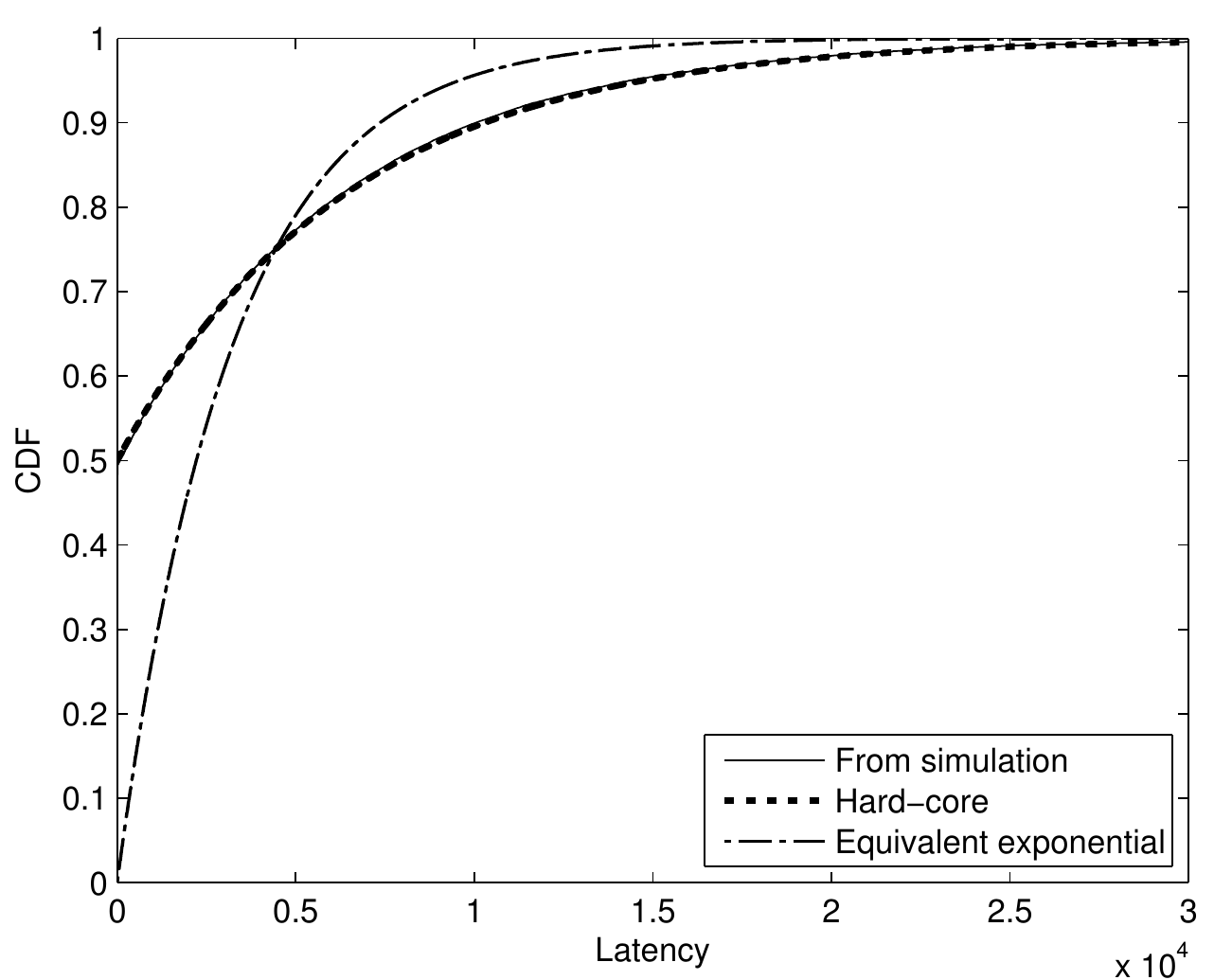}%
\caption{Cdf of latency for $N_f=1/32$.}%
\label{fig:cdf_latency_n32}%
\end{figure}

Figure \ref{fig:cdf_latency_n32} displays the latency distribution, using for comparison the hard--core distribution and the exponential distribution of parameter $W_o$.
One observes a close fit to the one proposed by the distribution function (\ref{eq:tdis-hardball}) of Conjecture \ref{conjecture3}: when a peer arrives, with probability one half, it disappears instantly; otherwise it follows an exponential distribution of average $2W_h$. In other words, not only the mean latency is much larger than in the fluid model (by a ratio $\frac{1}{N_f}$), but half of the peers will get a service time arbitrary larger compared to the other half (as $N_f$ goes towards $0$).

\begin{figure}%
\includegraphics[width=\columnwidth]{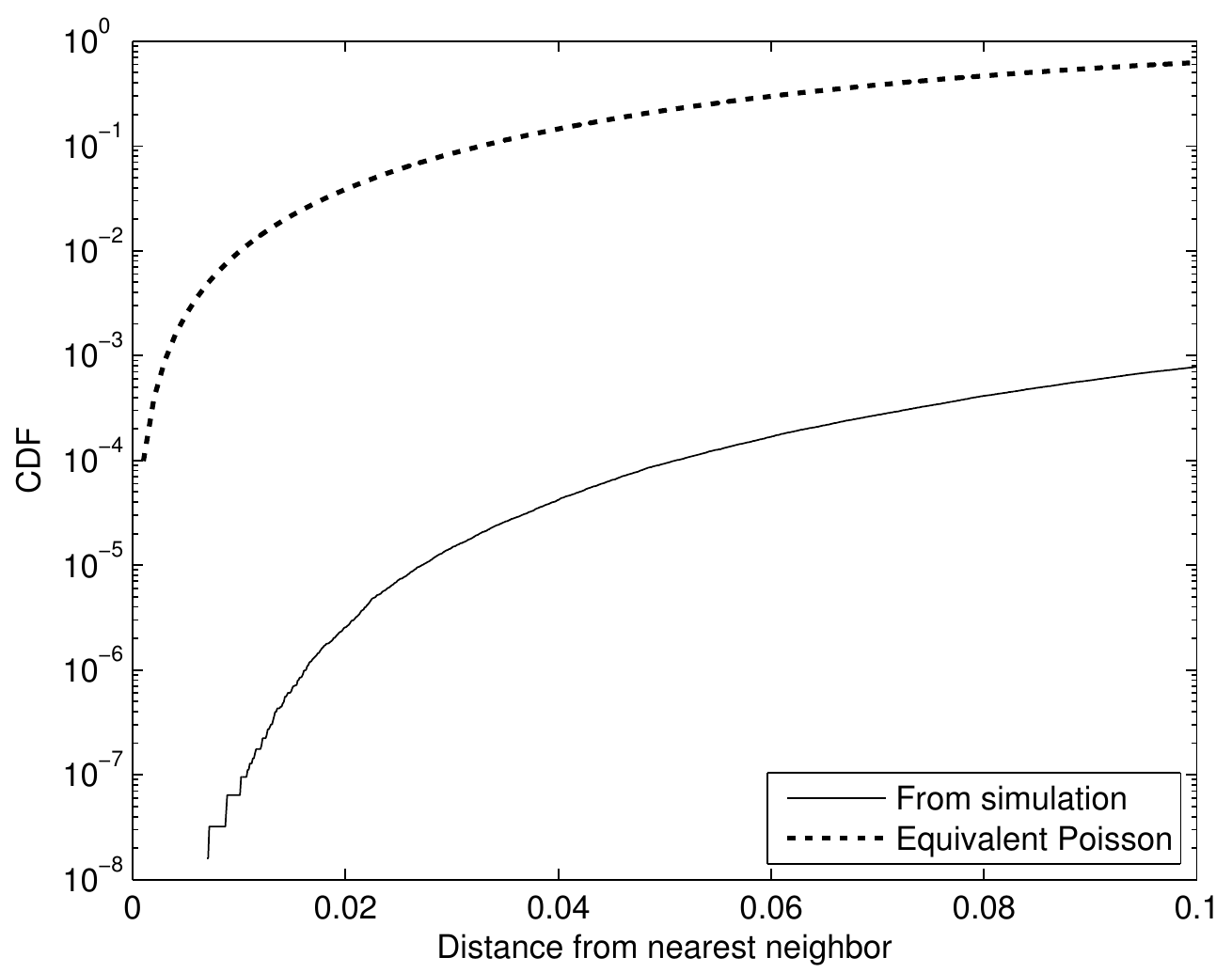}%
\caption{Cdf of nearest neighbor for $N_f=1/32$.}%
\label{fig:cdf_closest_n32}%
\end{figure}

The distribution of the closest neighbor is also of interest (cf. Figure \ref{fig:cdf_closest_n32}); the distribution has been truncated to the maximal distance $R$, as a peer does not ``see'' beyond $R$.

We see here the repulsion effect at its paroxysm: there are many orders of magnitude between the empirical distribution and the equivalent Poisson distribution. 
For instance, Poisson says that the probability to have at least one neighbor in range is $1-e^{-\lambda\pi W R^2}\approx 62.6\%$. In the stationary regime, this probability is only $0.078\%$, whereas the hard--core conjecture tells us that it will continue to decrease as $N_f$ goes to $0$.

\subsection{Intermediate Values}

We have no good formal description of the actual laws observed for intermediate values of $N_f$,  these distributions show a compromise  between the equivalent fluid and hard--core distributions.

\begin{figure}%
\includegraphics[width=\columnwidth]{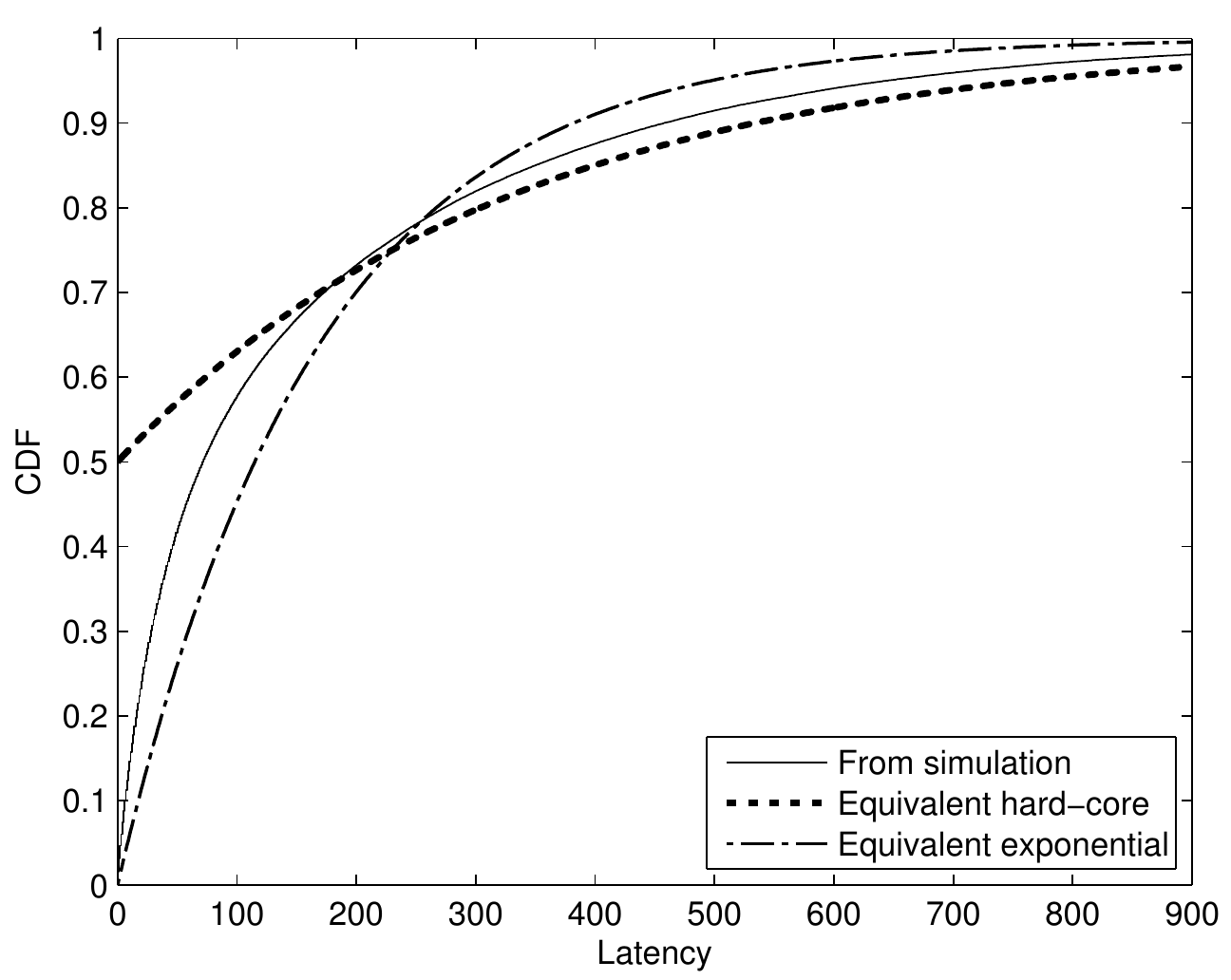}%
\caption{CDF of latency for $N_f=1$.}%
\label{fig:cdf_latency_n2}%
\end{figure}

\begin{figure}%
\includegraphics[width=\columnwidth]{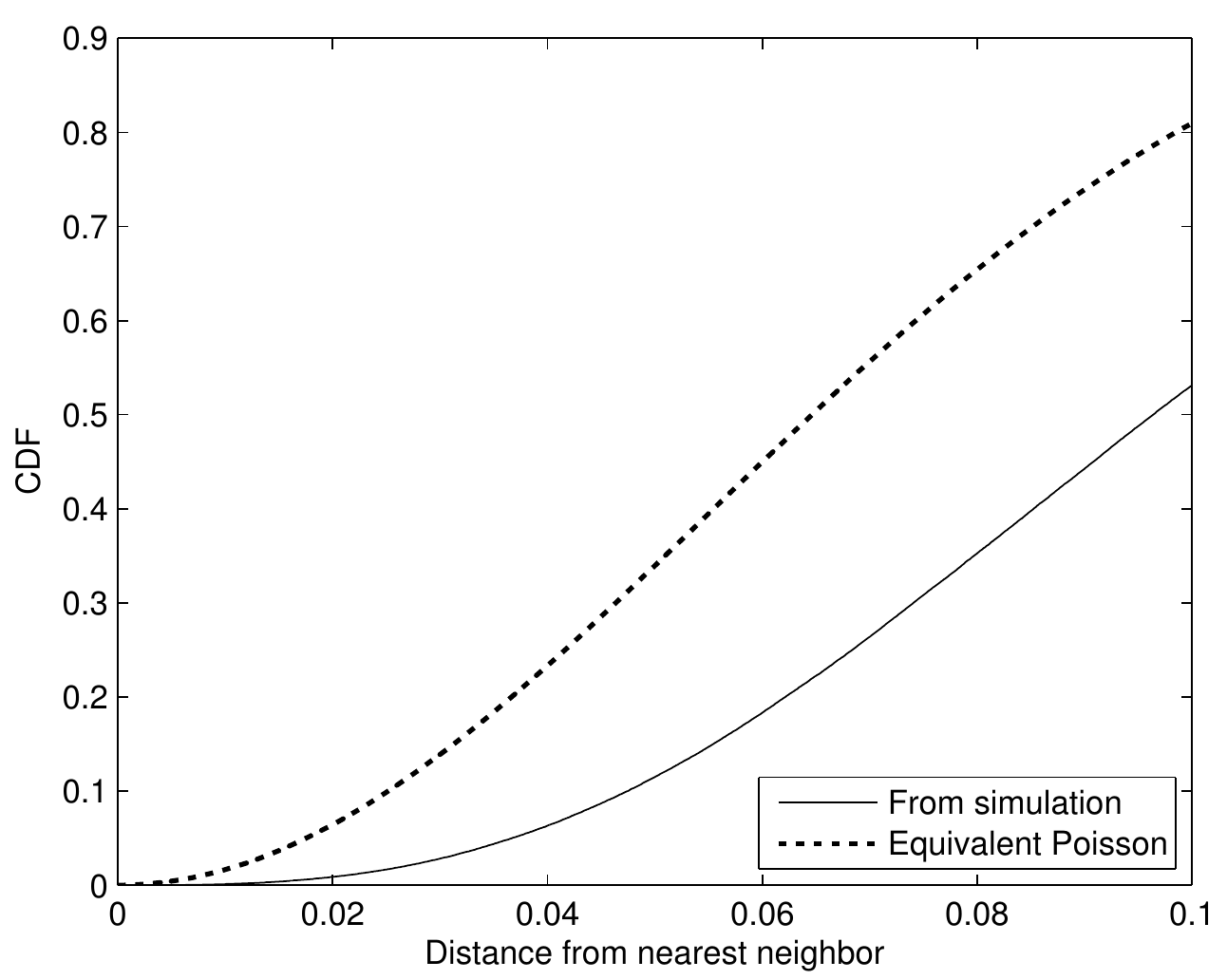}%
\caption{CDF of closest neighbor for $N_f=1$.}%
\label{fig:cdf_closest_n2}%
\end{figure}

In order to compare with the fluid and hard-core limits, we give the latency distribution (Figure \ref{fig:cdf_latency_n2}) and the closest neighbor distribution
(Figure \ref{fig:cdf_closest_n2}) for $N_f=1$.
One can see that these distributions show a compromise  between the equivalent fluid and hard--core distributions.

\subsection{Summary of Simulations}

For both the fluid and hard--core limits, simulations validate that we have a good description of the average system performance defined by $M$, but also of the latency distribution. For intermediate states, although the bounds still hold, it is better to rely on the heuristic, which gives quite accurate results on $M$, but with no details on the distribution.


\section{Network Capacity Constraints}
\label{sec:lim}

The aim of this section is to determine the capacity
required for the network elements in order to 
achieve the super-scalable regime identified above.

More precisely, so far, the only assumptions on the network
were that 1) the access is not a limitation anymore
(or not the only bottleneck);
2) the network is a bottleneck, resulting into a rate between
peers that depends on their distance and some range
or degree constraints. 

This section introduces an abstract network model
on which the P2P traffic will be mapped through 
some natural shortest path routing mechanism. We then
determine the mean {\em flow} that traverses a typical network element.
This flow of course depends on the protocols used in the network
which in turn determine the bit rate function.

For simplicity, we limit the study to the fluid limit of the system.

\subsection{Network Capacity Model}

We consider an underlying network made of routers and links between them.
A simple example is that where
\begin{itemize}
\item[(i)] routers form a Poisson point process of intensity $\theta$
in the plane;
\item[(ii)] links are the Delaunay edges (see e.g. \cite{FnT1}, Chapt. 4)
on this point process;
\item[(iii)] each peer is directly connected to the closest router
and the path between two routers is the shortest path (with minimal
hop count) on the Delaunay graph.
\end{itemize}

In this case, the number of links between two peers is
asymptotically proportional to the distance between them \cite{FnT1}.

For all straight lines of the plane, the point process of intersections
of this line with the edges forms a stationary point process
of intensity $\mu_e:=2\sqrt{\theta}$ on the line.
Denoting by $K$ the capacity 
of an edge, we get a total capacity per unit distance of
$\Xi:=\mu_e K$. 

Now, in order to simplify the evaluation of the P2P load on the underlying network,
we will assume that (a) $\theta$ is large enough so that the hop-count
between two peers can be seen as proportional to their distance
and the flow between them as a straight line;
(b) Any rate smaller than $\Xi l$ can be transported through a
segment of length $l$.\\

\noindent{\bf {Remark}}
In order to further justify the formula 
$f(r)= \frac C r 1_{r\le R}$ for the rate of two peers
at distance $r$ within the refined network model
presented above, one can use the bandwidth sharing 
formalism of \cite{BP}. A connection of Euclidean length $r$ uses
approximately $\gamma \frac r {\cal L}$ links 
where ${\cal L}=\frac{2}{3\sqrt{\theta}}$ 
is the mean length of a Delaunay edge of a Poisson 
point process of intensity $\theta$ (see \cite{SW08} p. 477) and 
where $\gamma$ is the (stretch) constant of the
shortest path algorithm (see \cite{FnT1}, Vol 2, Prop 20.7).
We assume that
each link is of capacity $K$. 
We consider the network as an open 
bandwidth sharing network \cite{BP} with 
connections of various classes arriving 
to the network, transferring a file of mean size
$F$ and leaving the network.
We write the bandwidth optimization problem in any given
state in this network as
\begin{eqnarray*}
&& \max \sum_{i} \log (\nu_i)\\
& & \mbox{under the constraints}\\
& & \sum_{i\in C_j} \nu_i \le K,
\end{eqnarray*}
where $\nu_i$ is the rate of connection $i$ and $C_j$
is the collection of connections that traverse link $j$ in this state.
Denoting by $\alpha_j$ the Lagrange multiplier 
associated with constraint $j$, we get that at the
optimum point, for all $i$
$$\frac 1 {\nu_i} -\sum_{j: i\in C_j} \alpha_j=0.$$
In the steady state regime (in both time and space),
the sequence $\alpha_j$ should be stationary and ergodic.
So, when denoting by $\alpha$ its mean, when $\theta$ is
large, $card\{j: i\in C_j\}$ is large too and
we get from spatial ergodicity that
if connection $i$ is of length $r$, namely uses
$\gamma r/\varepsilon_\rho$ links, then
$ \nu_i \approx \frac 1 \alpha \frac 1 {l(\nu_i)}$,
with $l(\nu_i)= card\{j: i\in C_j\}\approx \gamma \frac r{\cal L}$.
Hence
$\nu_i \approx {\cal L} \frac 1 {\alpha \gamma} \frac 1 r$
for $r\le R$.

\subsection{Flow Equations}

For the sake of easy exposition, we start with the model on the line.
The flow through the origin is
$$ \psi= 2 \E \sum_{X_i,X_j\in \Phi} f(|X_i-X_j|) 1_{X_i<0} 1_{X_j>0}.$$
In the fluid model, we can use the fact that the second
moment measure of $\Phi$ is $\beta^2$ times the Lebesgue
measure on $\R^2$ and Campbell's formula to get that
$$ \psi= 2 \int_{x<0}
\int_{y>0} f(y-x)\beta dx \beta dy = 2\beta^2 \int_{r>0} r f(r) dr.$$
The last expression comes from the change of variables
$r:=y-x,x:=x$.
Consider now the model on the plane.
Let $X_i=(x_i^1,x_i^2)$.

We make here the assumption that the bit flow between any two peers follows a straight line in the plane, and that the network capacity is defined by some constant $\Xi$, expressed in $bits.s^{-1}.m-1$, such that the maximal flow rate that can go through a segment.

Let $\Psi(\varepsilon)$ be the rate that goes through a segment $S$ of length $\varepsilon$. We can choose for instance 
$S=[(0,-\frac \varepsilon 2), (0,\frac \varepsilon 2)]$. 
Let $H^-$ denote the
left half-plane and $H^+$ the right half plane. Then $\Psi(\varepsilon)$
is
\begin{eqnarray*}
\Psi(\varepsilon) &= & 2 \E \sum_{\scriptsize
\begin{array}{c}
X_i\in \Phi\cap H^-,\\
X_j\in \Phi \cap H^+	
\end{array}
} f(|X_i-X_j|)
1_{[X_i,X_j]\cap S\ne \emptyset}\\
& = & 2 \iint_{\scriptsize
\begin{array}{c}
X\in  H^-,\\
Y\in  H^+	
\end{array}
}
f(|X-Y|) 1_{[X,Y]\cap S\ne \emptyset} \beta dX \beta dY
\\
& = &  2
\beta^2 
\iint_{\scriptsize
\begin{array}{c}
X\in  H^-,\\
Z\in  H^+	
\end{array}
}
f(|Z|) 
 1_{[X,Z+X]\cap S\ne \emptyset} dX dZ\\
& = &  4
\beta^2
\iint_{\scriptsize
\begin{array}{c}
r>0,\\
\theta \in [0,\frac{\pi}{2}]
\end{array}
}
f(r) r\sin(\theta) \varepsilon r dr d\theta
\\
& = &  4
\beta^2 \varepsilon
\int_{r>0} r^2 f(r)  dr
\end{eqnarray*}
where the third line comes from the change of variables
$Z:=Y-X$, $X:=X$.
So, by isotropy, the flow per unit length through any line of the plane is
$$\Psi=\Psi(1)= 4 \beta^2 \int_{r>0} r^2 f(r)  dr.$$

Using the fluid expression of the density
$$\beta=\beta_f= \sqrt{\frac{\lambda F}{2\pi\int_{r>0} r f(r) dr}}$$
we get the following key relation
\begin{equation}
\Psi=\Psi(1)= \frac 2 \pi \lambda F\frac {\int_{r>0} r^2 f(r)  dr}
{\int_{r>0} r f(r) dr}.
\end{equation}
In the TCP case
($f(r)= \frac C r 1_{r\le R}$),
we get 
\begin{equation}
\Psi_{\mbox{TCP}}=
2C \beta^2 \varepsilon R^2=
\frac 1 \pi \lambda F R.
\end{equation} 
In the UDP case
($f(r)=  C 1_{r\le R}$),
we get 
\begin{equation}
\Psi_{\mbox{UDP}}=
\frac 4 3 C \beta^2 \varepsilon R^3=
\frac 4 {3\pi} \lambda F R.
\end{equation} 

For the network to sustain the rate generated by our model, it is required that 
 \begin{equation}
\Psi\leq \Xi.
\label{eq:psixi}
\end{equation}
If one can assume, under some joint fluid limit,
that both the flow and the number of links going through
a segment are asymptotically deterministic,
then Condition \eqref{eq:psixi} is also sufficient for stability.
Studying the validity conditions of this hypothesis is, however,
beyond the scope of this paper.

Note that for both TCP and UDP, the condition \eqref{eq:psixi} does not depend on $C$. This surprising result means that in the fluid limit, we can arbitrarily scale the individual rate of connections (thus decreasing the latency) without changing the burden on the underlying network. Of course, there is a flaw in that reasoning, which is that increasing $C$ eventually impairs the validity of the fluid limit. In details, as $C$ increases, $N_f$ gets smaller so we tend to the hard-core limit where (i) there is unfairness as half of the peers get almost instant service compared to the other half; (ii) the average latency reaches an asymptotic value $\frac{1}{\lambda\pi R^2}$, so further increase of $C$ is meaningless.

\section{More General Rate Functions}
\label{sec:generalrate}

While we focused on TCP-like \eqref{eqr0} and UDP-like \eqref{eqr0U} functions, all our results can easily be generalized in the fluid limit to any rate function $f$ such that $\int_{r>0}rf(r)dr<\infty$. Even if $f$ has no maximal range $R$, we just have to replace $CR$ in \eqref{meanrate} by $\int_{r>0}rf(r)dr$ and proceed. This gives
\begin{equation}
\mu_f=\beta_f \gamma\text{, with }\gamma=2 \pi \int_{r>0}rf(r)dr\text{.}
\label{eq:generalrate}
\end{equation}
Once $\gamma$ is known, we can generalize \eqref{eq:cassimple} by
\begin{eqnarray}
\beta_f  =  \sqrt{ \frac{\lambda F}{\gamma}},\
\mu_f  =  \sqrt{ \lambda F \gamma},\
W_f   =  \sqrt{ \frac F{ \lambda \gamma}}.
\label{eq:generalperf}
\end{eqnarray}

Notice that the scaling in $\frac 1 {\sqrt{\lambda}}$ still holds.

Without a range $R$, $N_f$, which is $\pi R^2\beta_f$, is not properly defined, which impairs a direct introduction of $M$. However, if we have $\int_{r>0}r^2f(r)dr<\infty$, we can use
\begin{equation}
\tilde{R}:=\frac {\int_{r>0} r^2 f(r)  dr}
{\int_{r>0} r f(r) dr}o\end{equation}
instead of $R$ and extend the dimensional analysis accordingly
($\tilde{R}$ being interpreted as the \emph{typical} range of $f$).

Let us illustrate this method with a few concrete examples of type $f(r)=g(r)1_{r\le R}$.
\subsection{Affine RTT}
If $g$ is given by (\ref{eqr1}), then 
then the mean bit rate of a typical location of space is
$$\mu_f= \beta 2 \pi  \int_{r= 0}^R \frac C{r+q} r dr
=\beta_f 2\pi C\left( R-q\ln\left(1+\frac  Rq\right)\right),$$
so that we have
\begin{equation}
\label{meanrater1}
\gamma=  2 \pi  \int_{r= 0}^R \frac C{r+q} r dr=2\pi C\left( R-q\ln\left(1+\frac  Rq\right)\right).
\end{equation}
\subsection{Overhead}
For $g$ as in (\ref{eqr2}), after noticing the necessary condition $R\leq\frac Cc$ (each connection needs to use a minimal bandwidth $c$ for the overhead), we get
\begin{eqnarray*}
\mu_f  =  \beta_f 2 \pi  \int_{r= 0}^{R\wedge \frac C c} \left(\frac C{r} -c\right) r dr
\end{eqnarray*} 
so that
\begin{eqnarray}
\label{meanrater2}
\gamma =2 \pi  \int_{r= 0}^{R} \left(\frac C{r} -c\right) r dr=2\pi \left(RC-\frac{R^2c}{2}\right).
\end{eqnarray}
The best value for $R$ is $R=\frac Cc$, which gives $\gamma=\pi C^2/c.$

\subsection{Per Flow Rate Limitation}
The protocol or some physical constraints may limit the individual rates. If one assumes a maximal rate $U$ for each flow, we have $g(x)=(C/x)\wedge U$. This gives
\begin{equation}
\label{meanrateu}
\gamma =2 \pi  \int_{r= 0}^{R} \left(\frac Cr \wedge U \right) r dr
=
\left\{
\begin{array}{l}
\pi UR^2 \text{ if $C\ge UR$}\\
\pi \left(2CR-\frac{C^2} U\right)\text{ otherwise.}
\end{array}
\right.
\end{equation}
We find back \eqref{eq:cassimple} and \eqref{eq:cassimpleu} as special cases of \eqref{eq:generalperf} for $U\equiv \infty$ and $U\leq \frac{C}{R}$ (up to notation for the latter).

\subsection{SNR Wireless Model}
The setting is that where the bit rate function is
\begin{equation}
f(r) = \frac 1 2 \log\left(1+\frac C {r^\alpha}\right) 1_{r\le R}
\end{equation}
with $\alpha>2$ the path loss exponent, $C$ the Signal to Noise Ratio
at distance 1 and $R$ the transmission range.

In the case when $R$ is finite,
we will limit ourselves to the fluid case and to
the special case where $\alpha=4$ (the reason for the
las assumption being that
the relevant integral, namely
$\int_0^R \log\left(1+\frac C {r^\alpha}\right) r dr$,
can be then explicitly computed).
In this case, direct computations give that

\begin{equation}
\gamma = \pi \left( { {R^2}\log (1 + \frac C {R^4})
+ \sqrt{C} \arctan (\frac{R^2}{\sqrt{C}})} \right).
\end{equation}

The evaluation of the mean number of neighbors of a typical
node, namely $N_f=\pi R^2 \beta_f$,
allows one to identify the mean number of orthogonal channels
per unit space required to cope with the P2P load, namely

\begin{equation}
\beta_f N_f= \pi R^2 \frac{\lambda F}{\gamma}.
\end{equation}

In an infinite plane, this would require an infinite
number of orthogonal channels, which is of course not feasible.
However, it then makes sense to reuse spectrum in this
case, to the cost of an decrease of $C$ (resulting
from an increase of the noise power due to the presence
of distant interference).

In this sense, this scheme makes sense under appropriate
spectrum bandwidth assumptions, in the same way as the TCP
scheme makes sense under appropriate network capacity assumptions.

Notice that the integral
$\int_0^\infty \log\left(1+\frac C {r^\alpha}\right) r dr$
is finite. This allows us to consider the wireless
SNR model with an infinite range. In this case,
the result is much simpler:
for all $\alpha>2$,
\begin{equation}
\gamma = \frac{\pi^2 C^{\frac 2 \alpha}}{
2 \sin\left(\frac {2\pi}{\alpha}\right)}.
\end{equation}

\section{Extensions of the Basic Model}\label{sec:extend}
The aim of this section is to show that our analysis can be extended in several ways
and take important practical phenomena into account.
Unless otherwise stated, we will place ourselves in the fluid regime, but the dimensional analysis approach can be used with all extensions to relate the fluid limit to the real system through some function $M$. The only caveat is that if an extension introduces new parameters, $M$ can be a function of several dimensionless variables instead of $N_f$ only. This is illustrated by our first extension.

\subsection{Permanent Servers}

Assume that there exists some servers, or eternal 
seeders\footnote{This is distinct from the case where leechers can seed for some time after they complete their download, which is addressed in \ref{subsec:seeders}}. The motivation for considering this is for instance: (i) permanent servers can solve the issue of chunk availability by being able to provide any asked chunk; (ii) this allows one to consider hybrid systems which combine classical server solutions and a P2P approach; (iii) with our model, the latency goes to $\infty$ when $\lambda$ goes to $0$ (cf. \eqref{eq:hh}), which is not a desirable effect; servers or permanent seeders seem a perfect solution to prevent this.

We focus on the TCP case.

The servers are characterized by their density of bitrate $U_C$, expressed 
in $bit.s^-1.m^-2$, so that if $\beta_f$ is the peer density, 
a typical peer gets $\frac{U_C}{\beta_f}$ from the servers.

To describe the system, we need another dimensionless parameter in addition to $N_f$. We conveniently choose $\chi_C:=\frac{U_C}{\lambda F}$. $\chi_C$ expresses the ratio between the density of rate needed by the system and the density of rate provided by the servers. If $\chi_C\geq 1$, then the permanent rate from servers is sufficient to serve the peers, otherwise P2P is needed for stability.

Let us consider two limiting cases: the system is mainly client/server ($\chi_C\gg 1$), or the system is mainly P2P with a small server-assistance ($\chi_C\ll 1$). The case $\chi_C\ll 1$ can be seen as a scenario where servers are here mainly for insuring chunk availability.

If $\chi_C\gg 1$, then almost all resources come from the servers. We can deduce that the point process is hard--core (even if $N_f$ is large, if it is fixed and if $\chi_C$ grows, the servers can make newcomers leave before they have the occasion to reach another peer), so if a peer can collect all the available bandwidth in its range, the average latency will be
\begin{equation}
W_{C}\approx \frac{F}{\pi R^2 U_C}.
\label{eq:whc}
\end{equation}

For $\chi_C \ll 1$, we focus on the fluid limit ($N_f\gg 1$). Adapting \eqref{meanrate}, the rate of a peer is then
\begin{equation}
\mu_{f,C}=2\pi RC\beta_{f,C}+\frac{U_C}{\beta_{f,C}},
\end{equation}
from which we deduce 
\begin{equation}
W_{f,C} = \sqrt{ \frac {F-\frac{U_C}{\lambda}}{
\lambda 2 \pi CR}}=W_f\sqrt{1-\chi_C}\approx W_f.
\label{eq:wwithservers}
\end{equation}

Let us point out that the behavior of \eqref{eq:wwithservers} for $\chi_C$ close to 1 is not expected to be realistic, as the impact of the client/server behavior becomes prominent. For the hard--core process, one could also express $W_{h,C}$ as something that tends to $W_{h}$ if $\chi_C$ tends to $0$, which suggests that $M_C(N_f,\chi_C)$ admits a limit $M_C(N_f,0)=M(N_f)$ when $\chi_C$ tends to $0$. In words, the results presented in previous sections still hold if one assumes the existence of servers with relatively small bandwidth introduced to inject chunks into the system.

\vspace{-.1cm}
\subsection{Abandonment}
\vspace{-.1cm}
Here we consider the case where all leechers have some abandonment rate.
Let $a$ denote this rate. 
In the stationary state, we have $\lambda=(\frac{\mu_f}{F}+a)\beta_f$.
From \eqref{meanrate}, we deduce $\mu_f^2+\mu_faF=2\pi R C \lambda F$. The positive solution of this equation is 
\vspace{-.1cm}
\begin{equation}
\mu_f=\sqrt{2\pi R C \lambda F +\left(\frac{aF}{2}\right)^2}-\frac{aF}{2}\text{.}
\label{eq:abandonment}
\end{equation}
The analysis can hence be extended without difficulties. For instance, the abandonment ratio is given by $\frac{aF}{\mu_f+aF}$.

\subsection{Per Peer Rate Limitation} 
\vspace{-.1cm}
Due to the asymmetric nature of certain access networks (e.g. ADSL), the uplink rate is often the
most important access rate limitation. Let $U$ denote (here) the average upload capacity of a peer;
then the average rate in the fluid limit should be such that
\begin{equation}
\mu_f=\sqrt{\lambda F 2 \pi CR} \le U.
\label{eq:caspeeraccess}
\end{equation}
A natural dimensioning rule would then be to choose $R=\frac{U^2}{\lambda F 2 \pi C}$ in order to use all the available capacity.
\vspace{-.1cm}
\subsection{Leechers and Seeders}
\vspace{-.1cm}
\label{subsec:seeders}
When a leecher has obtained all its chunks, it can become a seeder and remains such 
for a duration $T_S$.
In this setting, there is a density of seeders $\lambda T_S$ in the stationary regime.

In the fluid limit with seeders, \eqref{meanrate} becomes
\begin{equation}
\label{meanratesee}
\mu_{f,S}=
(\beta_{f,S}+\lambda T_S) 2\pi C R.
\end{equation}
Using \eqref{eq:equil} and $F=W_{f,S}\mu_{f,S}$, we get
\begin{equation}
W_{f,S}^2+W_{f,S}T_S=W_{f}^2\text{.}
\label{eq:wfseeder}
\end{equation}
The positive solution of this equation is
\begin{equation}
W_{f,S}=\sqrt{W_{f}^2+\left(\frac{T_S}{2}\right)^2}-\frac{T_S}{2}.
\label{eq:wfseedersoluce}
\end{equation}
In particular, we have $W_{f,S}\approx W_{f}$ for $T_S\ll W_{f}$ and  $W_{f,S}\approx \frac{W^2_{f}}{T_S}$ for $T_s\gg W_{f}$.

By comparing \eqref{eq:wfseedersoluce} and \eqref{eq:abandonment}, one can interpret seeding as the exact opposite of abandonment: seeders, which improve the system, impact the latency the same way that abandonment, which degrades the system, impacts the rate.

We also remark that in a fluid model where rates are only determined by the upload access, we have (see \cite{benbadis08playing} for details)
\vspace{-.2cm}
\begin{equation}
W_{f,S}+T_S=W_{f}\text{.}
\label{eq:BCLclassic}
\end{equation}
We can see \eqref{eq:wfseeder} as the extension of \eqref{eq:BCLclassic} to the network-limited model.

At last, we propose to study the hard--core limit. Without seeder, a leecher can leave only if it finds a peer within range, and instant service happens with probability one half.
With seeders, a leecher is certain to complete its download if there is another peer in its neighborhood, as the latter will not leave the system before the former finishes. We can then notice that the configuration of peers (leechers and seeders) includes a spatial Poisson distribution of density $\lambda T_S$. In particular, the probability for a newcomer to find a peer within range $R$ is at least $1-e^{-\lambda  T_S \pi R^2}$. Therefore, for any $\epsilon>0$, if $T_S\geq \frac{-\log(\epsilon)}{\lambda \pi R^2}$, then leechers will get instant service with a probability greater than $1-\epsilon$.

This suggests that seeders may be a good antidote for systems where a hard--core behavior cannot be avoided: a seeding time of the same order of magnitude than the average latency in absence of seeders is enough to guarantee that most of the peers get instant download.

\subsection{Adaptive Range}

Consider the {\em constant number of nearest peers} model of 
Section \ref{sec:model}.
In the fluid limit, which can be reached by increasing $L$ until
it identifies to $N_f$, an approximate version of this model
is obtained by considering a range model with radius $R$ such that $R$,
the density $\beta$ and the target number of neighbors $L$ verify
\begin{equation}
\pi R^2 \beta =L.
\label{eq:pir2l}
\end{equation}
In this case, $\mu(x,\Phi)$ is
as in (\ref{eq:nonexp}) but with $R=\sqrt{\frac{L}{\pi \beta}}$.

In this section, we consider a general model with
$R=\kappa \beta^{-\alpha}$ with $\alpha$ a real parameter.  The
constant radius ball corresponds to the case $\alpha=0$ and the $L$
nearest neighbor case to $\alpha=\frac 1 2$.
Note that as $\beta$ depends on $R$, $R=\kappa \beta^{-\alpha}$
has to be seen as a fixed point equation for $\alpha>0$.

By dimensional analysis, one gets that for all $\alpha\neq \frac{1}{2}$,
all properties of the system only depend on the parameter
\begin{equation}
\rho=\frac{\lambda F}{C}\kappa^{\frac{3}{1-2\alpha}}.
\label{eq:rho}
\end{equation}
For $\alpha = \frac{1}{2}$ (nearest peers), the parameter is $\rho=\kappa$ (or equivalently $L$).

The fluid analysis gives $\mu_f  =  2\pi C\kappa \beta^{1-\alpha}$, so that
\begin{eqnarray}
\beta_f & = & \left(\frac {\lambda F}{2\pi C\kappa} \right)^{\frac 1 {2-\alpha}}\nonumber \\
W_f & = & \lambda^{-\frac{1-\alpha}{2-\alpha}} F^{\frac{1}{2-\alpha}} (2\pi C \kappa)^{-\frac{1}{2-\alpha}}\nonumber \\
\mu_f & = & (2\pi C\kappa)^{\frac 1{2-\alpha}} (\lambda F)^{\frac{1-\alpha}{2-\alpha}}.
\label{eq:casgen}
\end{eqnarray}
Notice that the algorithm which leads to this hence
consists in choosing a radius of the form
$ R=\kappa  \left(\frac {\lambda F}{2\pi C\kappa} \right)^{\frac \alpha {\alpha-2}}.$
For instance in the
{\em {constant number of nearest peers}} TCP case, we get
\begin{equation}
W_{{\mathrm{TCP}}}=\left(\frac{F}{2C}\right)^{\frac{2}{3}}\left(\frac{1}{\pi \lambda L}\right)^{\frac{1}{3}}.
\label{eq:TCPdegree}
\end{equation}

This is an interesting result: it means that in the fluid limit, TCP can achieve super--scalability even if each peer has a limited number of neighbors.

This is not the case for UDP, where the latency is $W_{{\mathrm{UDP}}}=\frac{F}{LC}$ (we still have scalability though).

We conclude this subsection by an asymptotic analysis
where all parameters are fixed but for $\lambda$ which tends to infinity.
We assume we are in the fluid regime (which
will lead to some restrictions on the set of parameters).

In view of (\ref{eq:casgen}), we will call $d=\frac 1 {2-\alpha}$ the {\em density exponent},
$l=\frac{\alpha-1}{2-\alpha}$ the {\em latency exponent} 
and $r= \alpha/(\alpha-2)$ the {\em radius exponent}.
We have the conservation rule
$d-l=1$, which is just a rephrasing of Little's law. Similarly
$N_f=K \lambda^{\frac {1-2\alpha} {2-\alpha}}$,
with $K$ a constant. So,
for $\lambda$ tending to $\infty$, the fluid regime requires that 
either $ \alpha > 2$ or $\alpha < \frac 1 2$.

Hence, there are 2 regimes when $\lambda\to \infty$:
\begin{itemize}
\item For $\alpha >2$, (which corresponds to $1< r < \infty$)
one gets at the same time 
$d<0$ and $l<0$, which means a peer density and a latency which both tend to 0 when $\lambda$ tends to $\infty$.
This is a rather surprising regime: the load per unit time and space tends to infinity; the density tends to 0 (there are no
peers around for delivering service); nevertheless, latency tends to 0 (i.e. when a peer
arrives, it is instantly served by invisible peers located at infinity). We will call
this regime {\em Heaven's--flash}.
\item For $\alpha<\frac 1 2$ (which corresponds to $-1/3<r<1$), one gets $d>0$ and $l<0$, which means
a peer density that tends to infinity and a latency which tends to zero when $\lambda$ tends to $\infty$.
This is the {\em swarm--flash} regime.
\end{itemize}
Notice the possible existence of a {\em critical--flash} regime, with $r=1$, $\alpha=\infty$, $d=0$ and $l=-1$,
where the density is a constant and the latency tends to 0. Another interesting though critical case is that where $\alpha=1/2$,
where the structural properties of the system do not depend on $\lambda$ anymore as shown by dimensional analysis.

\subsection{Mixed Extensions}
The proposed extensions, presented separately for sake of clarity,  can easily be interleaved, at least in the fluid limit. For instance, combining \eqref{eq:generalperf} and \eqref{eq:wfseedersoluce}, the average latency of a system with seeders and a rate function parameter $\gamma$ (cf \ref{sec:generalrate}) is
\begin{equation}
W_f=\sqrt{\frac{F}{\lambda \gamma}+\left(\frac{T_S}{2}\right)^2}-\frac{T_S}{2}\text{,}
\end{equation}

In order to illustrate the fact that the above
extensions are compatible, we analyze this
case in the setting where the uplink limitation is taken into account.

\begin{equation}
\label{meanratesee2}
\mu_f= (\beta_f+\lambda T_S) 2 \pi \int_{r= 0}^R (C/r)\wedge U r dr 
= (\beta_f+\lambda T_S) \xi(C,R,U).
\end{equation}

From Little's law applied to the leechers,
$ \beta_f= \lambda F/\mu_f$. Hence
$$ \beta_f(\beta_f+\lambda T_S) = \frac{\lambda F} {\xi(C,R,U)} .$$
The positive solution of this equation is
\begin{equation}
\beta_f= \frac{\lambda T_S} 2 \left(\sqrt{1+ \frac{4F}{\lambda T_S^2 \xi(C,R,U)}} -1 \right),
\end{equation}
which is an increasing function of $\lambda$.
Since $W_f=\beta_f/\lambda$,
\begin{equation}
W_f= \frac{T_S} 2 \left(\sqrt{1+ \frac{4F}{\lambda T_S^2 \xi(C,R,U)}} -1 \right).
\end{equation}
One can then mary this with the various ways of defining $R$ as a function
of $\lambda$.



\section{Conclusion}
\label{sec:conclusion}

The following general law quantifying
P2P super-scalability was identified:
in a P2P system with rate function
$g$ and range $R$, according to our model,
the stationary latency is of the form
\begin{equation}
\label{eq:grail}
W_o=M\left(\sqrt{\frac{\pi^2 R^4 \lambda F}{\gamma}}\right)
\sqrt{\frac{ F}{\lambda \gamma}},
\end{equation}
with $\gamma=2\pi\int_{0}^R g(r) dr$ and
with $M(x)$ a function which is larger than 1 and tends to 1 
when $x$ tends to infinity (if there is no range, \eqref{eq:grail} can still be used with the typical range $\tilde{R}$ defined in \ref{sec:generalrate}).

Both in the TCP case, i.e. for $g(r)=\frac C r$,
and in the UDP case, i.e. for $g(r)=C$,
the function $x\to M(x)$ is decreasing
(and has an explicit approximation).

With a decreasing $M$, Equation \eqref{eq:grail} exhibits two causes of super-scalability. First, there is the $\frac{1}{\sqrt{\lambda}}$ super--scalability that comes from the fluid term $W_f=\sqrt{\frac{ F}{\lambda \gamma}}$. This is the same type of super--scalability that was observed in the toy example. But there is also a super-scalability that comes from $M$, which expresses the surprising fact that increasing the arrival rate reduces the slow-down due to
the repulsion phenomenon identified in the paper. For $N_f$ large enough, the main cause of scalability is $W_f$, but otherwise, the effect of $M$ on super--scalability is not to be neglected.

The conditions for the super-scalability formula \eqref{eq:grail}
to hold were also identified:
First, the network should have the
capacity to cope with the P2P traffic. This translates
into the requirement
\begin{equation}
K \sqrt{\theta}>
\frac{2 \lambda F}{\gamma} \int_0^R r^2 g(r) dr, 
\end{equation}
where $\theta$ is the spatial intensity of routers
and $K$ the typical link capacity. In words, the linear
capacity of the network should scale like $\lambda$ if other parameters are unchanged.
Secondly the access should not be the bottleneck,
which translates into the requirement
\begin{equation}
U > \sqrt{\lambda F \gamma},
\end{equation}
where $U$ the (total) upload capacity of each peer.
In words, the latter should scale like the square root of $\lambda$.

We remark that the link capacity requirement is larger than the access requirement, which intuitively supports our initial motivation, which was that in future (wired) networks, the bottleneck should not be the access anymore.

Note that we are fully aware of the fact that, in the
the hard-core regime, our model might
fail due to the lack of adequate representation of the chunk level. We expect chunk availability to
become a crucial bottleneck in hard--core. So, if
$N_f:=\frac{R^4 \lambda F}{\gamma} \ll 1$, our conclusions
are probably overestimating the actual performance.

One of the future challenges in the research started by this paper is the extension to chunk-level modeling. Considering chunks leads to the issue of data availability, and a chunk-based system may be, in some scenarios, less stable that the models considered in this paper. For instance, a missing piece syndrome may be encountered in the form of growing spatial subpopulations missing at least one chunk. Parameters like the degree of altruism and the spatial intensity of permanent seeders can be expected to appear in the characterization of a stable regime.


\section{Appendix: Proof of Proposition 1 (Sketch)}
\label{sapp0}

Choose a number $z_0>0$ such that $f(z_0)>0$ and
split $D$ into cells with diameters at most $z_0$. Then all peers in a
cell with population higher than one receive service at least at rate
$f(z_0)$. It follows that the population of each cell is
stochastically dominated by an $M/M/\infty$ queue that is modified so
that a lone customer cannot leave. Since such queues are stable with
any input rate, the distribution of $(|\Phi_t|:\,t\ge0)$ is tight,
whatever the initial state $\Phi_0$. The ergodicity can now be shown by
a standard coupling argument: two realizations with different initial
states but same arrival process couple in finite
time.\hspace*{\fill}$\square$

\section{Appendix: Proof of Theorem 1}
\label{sapp1}

We work here on the torus $T$ of area $D$. Let $d$ denote the distance
on $T$ and $m$ the Haar measure.  Let $f:(0,\infty)\to(0,\infty)$ be a
positive function, and let $\Phi_t$ be the state of the SBD at time
$t$. For $x_0\in T$, let
$$ a=\int_{T}f(\|x-x_0\|) m({\D{x}}).$$
By translation invariance, $a$ is independent of the choice of
$x_0$. Further, the left hand side of the claim can be expressed as
\begin{equation}
  \label{thm1-lhs}
\E[ \sum_{x_i\in \Phi } f(||x_i||)]= \E(N_0)\frac{a}{D}.
\end{equation}
Consider now the P2P dynamics on $T$ in steady state.
For all $X\in \Phi_t$, let
\begin{eqnarray}
A_t(X) & = & \sum_{Y\in \Phi_t, Y\ne X}f(\|X-Y\|)\\
\A_t & = & \sum_{X\in \Phi_t} A_t(X),
\end{eqnarray}
where $A_t(X)$ is the death rate of point $X$ and $\A_t$ is the total
death rate of the SBD (here we assume that the mean file size $F$ is
equal to 1). The right hand side of the claim can be written as
\begin{equation}
  \label{thm1-rhs}
\E_0[ \sum_{x_i\in \phi \setminus\{0\}} f(||x_i||)]=\E_0(A_0(0)).
\end{equation}

By the rate conservation principle (e.g., \cite{baccellibremaud03},
1.3.3), applied to the stochastic process $N_t=\Phi_t(T)$, we get
\begin{eqnarray}
\label{cons1}
\lambda D= \E (\A_0)= \E (N_0) \E_0 (A_0(0)),
\end{eqnarray}
with $\E_0$ the (spatial) Palm probability of $\Phi_0$.
This relation says that the birth rate
$r^\uparrow=\lambda D$ should balance the death rate
$r^\downarrow=\E (\A)$. The relation
$\E (\A_0)= \E (N_0) \E_0 (A_0(0))$
follows from the definition of the Palm probability.

Let $\E^{\uparrow}$ denote the (time) Palm probability of the
SBD at birth epochs and $\E^{\downarrow}$ that at death epochs.
The rate conservation principle applied to the
stochastic process (total rate) $\A_t$, that we assume cadlag, gives
$$ r^{\uparrow} \E^\uparrow({\cal I}) = r^{\downarrow} \E^{\downarrow}({\cal D})$$
with ${\cal I}=\A_{0+}-\A_0$ the total rate increase and
${\cal D}=\A_0-\A_{0+}$ the (absolute value of the) total rate decrease. 
Since $ r^{\uparrow}= r^{\downarrow}$, we get that
$$ \E^{\uparrow}({\cal I}) = \E^{\downarrow}({\cal D}).$$

From the PASTA property \cite{baccellibremaud03}, and the fact that 
births are uniform on $T$,
$$\E^{\uparrow}({\cal I})= 2 \E(N_0) \frac a D .$$
 
The (total) death point process admits a stochastic intensity
w.r.t. the filtration ${\cal F}_t=\sigma(\Phi_s,s\le t)$ equal to
$\A_t$.  Hence, it follows from Papangelou's theorem (e.g.,
\cite{baccellibremaud03}, Theorem 1.9.2) that 
$$ \frac {d {\mathbb P}^{\downarrow}}{ d{\mathbb P} } \mid_{{\cal F}_{0-}} =
\frac{\A_0}{\E(\A_0)}.$$
Since the decrease (in state $\Phi_{0-}$) is
of magnitude $A_0(X)$ (w.r.t. $\Phi_{0-}$) with probability $\frac
{A_0(X)}{\A_0}$ (w.r.t. $\Phi_{0-}$), we get
\begin{eqnarray*}
\E^{\downarrow}({\cal D}) & = &
2 \E \left( \frac {\A_0}{\E(\A_0)}
\sum\limits_{X\in \Phi_0} \frac {A_0(X)}{\A_0} A_0(X) \right) \\
& = &
2\frac{ \E \left(\sum\limits_{X\in \Phi_0} (A_0(X))^2 \right) }
{\E \left(\sum\limits_{X\in \Phi_0} A_0(X) \right)}.
\end{eqnarray*}
Hence, when using the fact that
$$
\frac{ \E \left(\sum\limits_{X\in \Phi_0} (A_0(X))^2 \right) }
{\E \left(\sum\limits_{X\in \Phi_0} A_0(X) \right)}
=
\frac{ \E_0 \left((A_0(0))^2\right)}
{ \E_0 \left(A_0(0)\right)} ,$$
the rate conservation principle for total rate gives:
\begin{eqnarray}
\label{cons2}
\E(N_0) \frac a D=
\frac{ \E_0 \left((A_0(0))^2\right)}
{ \E_0 \left(A_0(0)\right)} .
\end{eqnarray}

Recalling (\ref{thm1-lhs}) and (\ref{thm1-rhs}), we now note that
Theorem \ref{conjecture1} follows from the fact that 
$\E_0\left((A_0(0))^2\right) \ge \E_0 \left(A_0(0)\right)^2.
$

\section{Appendix: Sketch of Proof of Theorem 2}
\label{sapp2}

Assume for simplicity that $f$ is bounded. We proceed as in the fluid
limit of a queue, by scaling the arrival and service rates
appropriately, and consider a sequence of systems indexed by $n$,
where $n$ is a parameter that tends to infinity. Our assumption is
that the arrival rate in system $n$ is $\lambda_n=\lambda n$, and the
mean file size in system $n$ is $F_n=F n$. 

We tessellate the plane with a grid made of squares of side $\delta$,
and time with a grid of width $\eta$. Hence, the mean number of
arrivals in a typical square and a typical time interval is $\lambda n
\delta^2 \eta$ for all $n$. In addition, the strong law of large numbers
(SLLN) shows that the
random number of arrivals $A^t_n$ in a typical square in the time
interval $(t,t+\eta)$ is such that $A_n^t/n$ tends a.s.\ to the
constant $\lambda \eta \delta^2$ when $n$ tends to infinity.

The next task is to show that the number of peers $N_{n}^t (k,l)$
present at time $t$ in the square with
coordinates $(k\delta,l\delta)$ is such that
$N_{n}^t (k,l)/n$ converges a.s. to some deterministic
limit $\beta^t(k,l)\delta^2$. We then get that the
number of deaths in this square in the time interval $(t,t+\eta)$,
denoted by $D^t_n(k,l)$, satisfies
\begin{eqnarray*}
&&\lim_{n\to \infty} \frac 1 n D^t_n(k,l)\\
&=& \frac 1 F \beta^t(k,l) 
\sum_{p,q}  f[(p\delta,q\delta),(k\delta,l\delta)] \beta^t(p,q) 
\delta^4 \eta.
\end{eqnarray*} 
This follows from the fact that the probability that a typical peer in
the square $(k\delta,l\delta)$ dies approximately with probability
$$\frac {\eta} {F n}
\sum_{p,q}  f[(p\delta,q\delta),(k\delta,l\delta)]  N_{n}^t (p,q)
\text{ that tends to
}$$
$$\frac 1 F
\sum_{p,q}  f[(p\delta,q\delta),(k\delta,l\delta)] \beta^t(p,q) 
\delta^2 \eta,
$$ 
so that the number of deaths tends
to the announced limit. (Notice however that this discretization
does not make sense for, e.g., $f(x,y)=C/|x-y|$, as 
$f[(k\delta,l\delta),(k\delta,l\delta)]=\infty$.)

Hence, by letting $\delta$ and $\eta$ tend to 0, 
we get that the function $\beta^t(x)$ which is the
value of the density at $x\in \R^2$ at time $t$ in the fluid
regime satisfies
the differential equation
\begin{equation}
 \label{eq:kf}
\frac d {dt}
\beta^t(x) = \lambda - \frac {\beta^t(x)} F \int_{\R^2} f(x,y) \beta^t(y) dy.
\end{equation}
The steady state of this is
$$ \frac {\lambda }{\beta(x)} = \frac 1 F \int_{\R^2} f(x,y) \beta(y) dy.$$
A translation invariant solution of this is
$$\beta^2= \frac {\lambda F}{  \int_{\R^2} f(x,y) dy},$$
which is the ``fluid solution''.

\section{Appendix: Justification of the Heuristic}
\label{rsappendheur}

In order to derive the heuristic of Section \ref{secheur},
we use the balance equation for the second order factorial moment density, which reads
\begin{eqnarray}
\label{eqo2}
& &\hspace{-.5cm} 2\beta_o \lambda  =
 2 m_{[2]}(x,y)  
\frac C F \frac {1_{||x-y||\le R}} {||x-y||}\\
& & + \frac C F
\int_D m_{[3]}(x,y,z)
\left(
\frac {1_{||x-z||\le R}} {||x-z||}
+\frac {1_{||y-z||\le R}} {||y-z||} 
\right)
dz ,
\nonumber
\end{eqnarray}
for all $x$ and $y$.
We then use the following approximations:
\begin{eqnarray*}
 m_{[3]}(x,y,z) & \approx & \frac{ m_{[2]}(x,y) m_{[2]}(x,z)} {\beta_o}\\
 m_{[3]}(x,y,z) & \approx & \frac{ m_{[2]}(x,y) m_{[2]}(y,z)} {\beta_o}.
\end{eqnarray*}
Then, we get from (\ref{eqo2}) that
\begin{eqnarray*}
& &\hspace{-.5cm} \beta_o \lambda \approx
m_{[2]}(x,y)  
\frac C F \frac {1_{||x-y||\le R}} {||x-y||} \\ & & +
m_{[2]}(x,y)  \frac C F \frac 1 2 \int_D 
\frac {1_{||x-z||\le R}} {||x-z||} \frac{m_{[2]}(x,z)}{\beta_o} dz\\
& & +
m_{[2]}(x,y) \frac C F \frac 1 2 \int_D 
\frac {1_{||y-z||\le R}} {||y-z||}  \frac{m_{[2]}(y,z)}{\beta_o}
dz,
\end{eqnarray*}
that is
\begin{equation}
m_{[2]}(x,y) \approx \lambda F \frac{\beta_o}{\frac {C 1_{||x-y||\le R}} {||x-y||}+ \mu_o}.
\end{equation}
with
$\mu_o   = :  C \int_{B(0,R)} \frac{m_{[2]}(0,z)}{\beta_o} \frac 1{||z||} dz $.
So
\begin{equation}
\begin{array}{rcl}	
\mu_o 
& \approx & \lambda F 2\pi C \int_0^R \frac 1 {\mu_o +\frac C r} dr\\
& = &  \lambda F 2\pi C  \left(\frac R {\mu_o} -\frac C {\mu_o ^2} \ln(1 + \frac{\mu_o R} C)\right),
\end{array}
\end{equation}
which is our departure point in Section \ref{secheur}.

\bibliographystyle{IEEEtran}

\label{LastPage}

\end{document}